\documentclass[aps,prb,twocolumn,superscriptaddress,showpacs,notitlepage,nolinenumbers]{revtex4-2}
\usepackage[english]{babel}
\usepackage[export]{adjustbox}

\usepackage{xcolor} 
\usepackage{graphicx} 

\usepackage{amsmath,amsfonts,amssymb}
\usepackage{mathtools} 
\usepackage{bm}

\newcommand{\txup}[1]{\textsuperscript{#1}}

\newcommand{\tbf}[1]{\textbf{#1}}

\newcommand{\bcol} {\textcolor{blue}}

\newcommand{\opr}[1]{\hat{#1}}
\newcommand{\id}{\opr{1}}

\newcommand{\idm}{\mathbb{I}}
\newcommand{\ii}{\iota}
\newcommand{\e}{\,\mathrm{e}}

\newcommand{\ve}[1]{{\boldsymbol{#1}}}
\newcommand{\tx}[1]{{\text{#1}}}

\newcommand{\del}{\partial}
\newcommand{\no}{\nonumber}
\newcommand{\ro}{\opr{\rho}}

\newcommand{\mesr}[1]{\mathcal{D}\hspace{-2pt}\left[#1\right]}

\newcommand{\partfn}{\mathcal{Z}}

\newcommand{\beq}{\begin{equation}}
  \newcommand{\eeq}{\end{equation}}
\newcommand{\bqa}{\begin{eqnarray}}
\newcommand{\eqa}{\end{eqnarray}}

\newcommand{\et}{\eta}
\newcommand{\ebar}{\bar{\eta}}
\newcommand{\ze}{\zeta}
\newcommand{\zbar}{\bar{\zeta}}

\newcommand{\pbar}{\bar{\psi}}

\newcommand{\cd}{\opr{c}^{\dagger}}
\newcommand{\cc}{\opr{c}^{\vphantom{\dagger}}}

\usepackage{physics} 
\usepackage{easyReview} 

\usepackage[pdftex]{hyperref}
\hypersetup{colorlinks=true,citecolor=magenta,linkcolor=blue,breaklinks=true}

\usepackage{calc}
\usepackage{natbib} 

\usepackage{booktabs} 
\usepackage{array} 

\usepackage{orcidlink}
\graphicspath{{./Pictures/}}

\newcommand{\actn}{\mathcal{S}} 
\newcommand{\Sent}{\actn_{\tx{ent}}} 
\newcommand{\Sint}{\actn_{\tx{int}}} 
\newcommand{\PP}{\mathrm{P}} 

\begin{document}

\begin{abstract}
We provide a prescription to construct R\'{e}nyi and von Neumann
entropy of a system of interacting fermions from a knowledge of
its correlation functions. We show that R\'{e}nyi entanglement entropy of interacting fermions in arbitrary
dimensions can be represented by a Schwinger Keldysh free energy on
replicated manifolds with a current between the replicas. The current
is local in real space and is present only in the subsystem which is
not integrated out. This allows us to construct a diagrammatic
representation of entanglement entropy in terms of connected
correlators in the \emph{standard field theory with no replicas}. This
construction is agnostic to how the correlators are calculated, and one
can use calculated, simulated or measured values of the correlators in
this formula. Using this
diagrammatic representation, one can decompose entanglement into
contributions which depend on the one-particle correlator, two
particle correlator and so on. We provide analytic formula for the
one-particle contribution and a diagrammatic construction for higher
order contributions. We show how this construction can be extended for
von-Neumann entropy through analytic continuation.  For a practical implementation of a
quantum state, where one usually has information only about few-particle
correlators, this provides an approximate way of calculating
entanglement commensurate with the limited knowledge about the underlying
quantum state.
\end{abstract}
\title{Building Entanglement Entropy out of Correlation Functions for Interacting Fermions}
\author{Saranyo Moitra\,\orcidlink{0000-0001-7912-1961}}\email{smoitra@theory.tifr.res.in}
 \affiliation{Department of Theoretical Physics, Tata Institute of Fundamental
 Research, Mumbai 400005, India.}
\author{ Rajdeep Sensarma\,\orcidlink{0000-0002-2136-5354}}\email{sensarma@theory.tifr.res.in}
\affiliation{Department of Theoretical Physics, Tata Institute of Fundamental
 Research, Mumbai 400005, India.}

\date{\today}

\maketitle
\section{Introduction}

A quantum many body state encodes non-local correlations between degrees
of freedom in one part of the system with those in another part;
i.e. the information about the state is distributed amongst degrees of
freedom which are typically far away from each other. A
simple way to see this is to expand a quantum many body state in a basis which is
tensor product of local basis states. The complex quantum amplitudes of this
expansion (or the many-body wavefunction) store the information about these
non-local correlations. The
most obvious example of this is quantum statistics: e.g. Fermions
cannot share a quantum state, which can impose long range non-local
constraints on wavefunctions of many Fermions.

If one is interested in observables which have support only in the
Hilbert space of a subsystem $A$, one can trace over the degrees of
freedom in the complimentary subsystem $B$ and construct a reduced density
matrix (RDM) $\ro_A$~\cite{vonneumann1932mathematische}. This density operator will reproduce all observables
 in the subsystem and has the usual interpretation of a ``classical'' ensemble 
 of quantum states as specified by its spectral decomposition. 
If the RDM represents a pure quantum state in the subsystem, the
original state is separable, otherwise it is entangled~\cite{HorodeckiReview}.

Entanglement and its related measures have played an important role in
wide ranging fields including 
quantum information and computation
\cite{Bennett.DiVincenzo_N00_QuantumInformationComputation,nielsen2010quantum}, 
foundations of quantum mechanics
\cite{EPR,*BELL_RMP66_ProblemHiddenVariables,*Clauser.Holt_PRL69_ProposedExperimentTest,*Zeilinger_RMP99_ExperimentFoundationsQuantum}, 
black-hole physics and quantum gravity
\cite{ryu_Phys.Rev.Lett.2006_HolographicDerivationEntanglement,*hayden_J.HighEnergyPhys.2007_BlackHolesMirrors,*susskind_EREPR,*penington_J.HighEnerg.Phys.2020_EntanglementWedgeReconstruction,*Das:2020xoa}, 
and quantum condensed matter systems
\cite{Amico.Vedral_RMP08_EntanglementManybodySystems,laflorencie_PhysicsReports2016_QuantumEntanglementCondensed}. 
An important measure of entanglement is the bipartite entanglement entropy (EE), 
which is the classical entropy corresponding to the probability distribution 
specified by the eigenvalues of $\ro_A$. 
In quantum many body systems, the scaling of EE with subsystem size is used to 
fingerprint states
~\cite{Eisert.Plenio_RMP10_ColloquiumAreaLaws,
	laflorencie_PhysicsReports2016_QuantumEntanglementCondensed,
	Holzhey.Wilczek_NPB94_GeometricRenormalizedEntropy,
	*Calabrese_2004,
	*Calabrese_2009,
	gioev_Phys.Rev.Lett.2006_EntanglementEntropyFermions,
	casini_J.Phys.A:Math.Theor.2009_EntanglementEntropyFree}, 
detect presence of topological order
~\cite{kitaev_Phys.Rev.Lett.2006_TopologicalEntanglementEntropy,
	*levin_Phys.Rev.Lett.2006_DetectingTopologicalOrder,
	*jiang_NaturePhys2012_IdentifyingTopologicalOrder} 
or occurrence of  quantum phase transitions
~\cite{Vidal.Kitaev_PRL03_EntanglementQuantumCritical,*SubirON,*SubirWitczakKrempa,*ju_Phys.Rev.B2012_EntanglementScalingTwodimensional}. 
In fact, exotic states like spin liquids\cite{Savary.Balents_RPP16_QuantumSpinLiquids} are often
best described by the ``entanglement patterns'' in the ground state
~\cite{zhang_Phys.Rev.Lett.2011_EntanglementEntropyCritical,*isakov_NaturePhys2011_TopologicalEntanglementEntropy, *pretko_Phys.Rev.B2016_EntanglementEntropyQuantum}. 
More recently the entanglement scaling of excited states in
the middle of the spectrum~\cite{Abanin_EntanglementMBL} has been used to classify whether a
disordered interacting system
is in an ergodic phase, where local observables can be described by
usual statistical mechanics, or in a many body localized phase~\cite{nandkishore_AnnRevCondens.MatterPhys2015_ManyBodyLocalization}, where
laws of statistical mechanics fail to apply. 
The question of thermalization~\cite{DAlessio.Rigol_AP16_QuantumChaosEigenstate} in a system has also been tracked through the time evolution 
of entanglement under non-equilibrium dynamics ~\cite{calabrese_J.Stat.Mech.2005_EvolutionEntanglementEntropy,Chakraborty.Sensarma_PRL21_NonequilibriumDynamicsRenyi,moitra_entanglement_2020}. Thermalizing systems show a linear growth of
entanglement, while many body localized systems have a slower
logarithmic growth of entanglement~\cite{Chiara.Fazio_JSM06_EntanglementEntropyDynamics,Bardarson.Moore_PRL12_UnboundedGrowthEntanglement}. The growth of entanglement has
also played a crucial role in the development of ideas related to
quantum chaos~\cite{maldacena_J.HighEnerg.Phys.2016_BoundChaos,*Swingle.Hayden_PRA16_MeasuringScramblingQuantum,Mi.Chen_S21_InformationScramblingQuantum}.
Experimental measurements of EE in many body systems have been performed on a variety of platforms including ion traps~\cite{Islam.Greiner_N15_MeasuringEntanglementEntropy,Joshi.Zoller_23_ExploringLargeScaleEntanglement}, ultracold Rydberg atoms~\cite{Lukin.Greiner_S19_ProbingEntanglementManybody,Tajik.Schmiedmayer_NP23_VerificationAreaLaw}, and superconducting qubits~\cite{Karamlou.Oliver_23_ProbingEntanglementEnergy}.

There are only a few methods to calculate the EE in
interacting many body systems, either in thermal or in the ground
state. Even fewer methods can tackle non-equilibrium dynamics of EE. 
The most direct method is to obtain the quantum state by exact
diagonalization, calculate the reduced density matrix and hence the
EE~\cite{Kim.Huse_PRL13_BallisticSpreadingEntanglement,Laflorencie.Affleck_PRL06_BoundaryEffectsCritical}. While this is the most widely used method, it is
limited to small finite size systems in one dimension, since the
specification of a quantum many body state requires knowledge of an
exponentially large (in system size) number of quantum amplitudes. 
	Quantum Monte Carlo based methods have been extensively used for systems in one and two dimensions
	~\cite{
		Hastings.Melko_PRL10_MeasuringRenyiEntanglement,
		*Grover_PRL13_EntanglementInteractingFermions,
		*McMinis.Tubman_PRB13_RenyiEntropyInteracting,
		*Drut.Porter_PRB15_HybridMonteCarlo}.  
In the cases where EE is expected to have a weak scaling
with system size, numerical techniques using Density Matrix Renormalisation Group ideas~\cite{White_PRL92_DensityMatrixFormulation,*Schollwock_AoP11_DensitymatrixRenormalizationGroup} in one dimension and
tensor network based methods~\cite{Vidal_PRL07_EntanglementRenormalization,*Evenbly.Vidal_JSP11_TensorNetworkStates} in higher dimensions have been used to calculate EE. For non-interacting systems or integrable systems, progress
can be made using specialized numerical techniques, as one can reduce
the complexity of the calculation~\cite{Peschel_2003,*Peschel_2009}. Standard field theoretic techniques
for calculation of entanglement
entropy~\cite{Calabrese_2004,*Calabrese_2009, casini_J.Phys.A:Math.Theor.2009_EntanglementEntropyFree} use
replica methods with complicated boundary conditions. This limits
their scope as it is often impossible to obtain solutions of even
simple problems with the complicated boundary conditions. For 1+1
dimensional CFTs~\cite{Holzhey.Wilczek_NPB94_GeometricRenormalizedEntropy,*Calabrese_2004}, where correlators are tightly constrained, one can
obtain exact analytical answers for leading and subleading scaling of EE. 

A key problem is calculating EE of a state is the
following: we have an operational prescription to calculate
EE if we know the exact quantum state; however for a
generic system, there is no such prescription to calculate the
EE in terms of the correlation functions. There are
two aspects to this issue: (a) A prescription to obtain entanglement
from correlators can lead to efficient estimates of entanglement in
numerics, since efficient algorithms 
to calculate correlation functions
already exist in the literature~\cite{Becca.Sorella_17_QuantumMonteCarlo,Schollwock_AoP11_DensitymatrixRenormalizationGroup}. 
There are also a large class of
analytic approximation schemes~\cite{Mahan_00_ManyParticlePhysics,Altland.Simons_10_CondensedMatterField} for calculating interacting correlation functions, which have been
developed over the years. With a prescription connecting correlations
to EE, these methods can increase the scope of study
of entanglement in large quantum systems. 
We note that in the special cases where this prescription is
known, like in non-interacting systems~\cite{Peschel_2009,gioev_Phys.Rev.Lett.2006_EntanglementEntropyFermions} or 1+1D CFT~\cite{Calabrese_2009}, our knowledge of EE is vastly more advanced
than the cases where such a prescription is not known. (b) In any
realistic situation, it is impossible to
know the exact quantum state; however there are experimental probes to obtain information about the
correlation functions~\cite{Schweigler.Schmiedmayer_N17_ExperimentalCharacterizationQuantum}. Even in highly controllable systems like
ultracold atoms~\cite{Gross.Bloch_S17_QuantumSimulationsUltracold}, experiments can at best have knowledge of few-body
correlations~\cite{Preiss.Jochim_PRL19_HighContrastInterferenceUltracold,Folling.Bloch_N05_SpatialQuantumNoise,Hart.Hulet_N15_ObservationAntiferromagneticCorrelations}. A prescription to calculate entanglement
entropy from knowledge of correlation functions would thus not only
enhance the theoretical space for such calculations, it will be the only consistent
description of realistic experimental situations. Here we would like
to note that if one requires the knowledge of all $m$-body correlators
in a system to calculate EE, the complexity of the problem
is same as knowing the full quantum state. Thus it would be useful to
have estimates of entanglement which involve only few body correlators,
and one should be able to improve these estimates if information about
higher order correlators become available.

In this paper, we take on the task of constructing an operational
prescription for computing EE of a generic
interacting Fermionic system in terms
of its correlation functions. We consider the system to be made of
mutually exclusive regions $A$ and $B$, with a Hilbert space which
is a tensor product of Hilbert spaces for degrees of freedom in $A$
and $B$. We are interested in the EE of the system
when the degrees of freedom in $B$ are traced out. We use Schwinger Keldysh field theory~\cite{Kamenev_11_FieldTheoryNonEquilibrium,Rammer_07_QuantumFieldTheory},
which allows us to consider ground states, thermal systems, and closed and open quantum systems
evolving in time out of thermal equilibrium on the same footing. Our
prescription is agnostic to these different situations. (i)  We show
that the $n$\txup{th} R\'{e}nyi entropy, $S^{(n)}$, of a system of interacting Fermions is the
Schwinger Keldysh free energy of a system of $n$ replicas with
``inter-replica currents'' flowing only in the subsystem $A$. These
currents are local in space (i.e. between same lattice
sites, or same location) and in time (currents are present only at the
time of measurement). 
The matching of fields across replica-s in standard field theory technique for calculating EE is replaced by the \emph{inter-replica currents} in our formalism. 
We argue that the doubling of fields in the SK
field theory is a crucial ingredient which allows boundary conditions
to be replaced by quadratic current-like terms, and we do not know of
any method 
{which can achieve this} in a single contour Lorentzian or
Euclidean field theory. (ii) Using this identification of EE
with a free energy, we provide a prescription to calculate
it in terms of correlations in a single replica theory,
which does not involve complicated boundary conditions (it has the
same boundary conditions as the usual field theory). (iii) We show that if one only has
information of up to $m$-particle connected correlators, one can construct an
estimate of EE, which can be improved if information about
$m+1$-particle correlators become available. More precisely, for the $n$\txup{th} order 
R\'{e}nyi Entanglement Entropy (REE), $S^{(n)}$, we show
\begin{equation}
S^{(n)}=S^{(n)}_{1\PP}+S^{(n)}_{2\PP}+S^{(n)}_{3\PP}+\cdots,
\end{equation}
where $S^{(n)}_{m\PP}$ involves up to $m$-particle connected correlators, $G^{(m)}_c$,
and does not involve any correlators of higher order (particle number). 
For $m>1$, $G^{(m)}_c$ is simply related to
the connected part of the expectation value of a normal ordered string of Fermion operators ($\cd,\cc$) w.r.to the state,
\begin{equation}\label{eq:Gmc.operators}
		G^{(m)}_c\sim%
		\ev*{\,\underbrace{\cd\cd\cdots\cd}_{m \rm\ times}\,\underbrace{\cc\,\cc\,\cdots\cc}_{m \rm\ times}\,}_c.
\end{equation}
We calculate $S^{(n)}_{1\PP}$ exactly and provide a prescription of Feynman diagrams
for calculating $S^{(n)}_{m\PP}$ for $m>1$ in terms of higher order
connected correlators. This decomposition
would be most useful when the state of the system is close to that of
a Gaussian theory (this could be a symmetry broken mean field theory)
and higher order connected correlators are parametrically
smaller. (iv) The von-Neumann entropy, $S$, does not have a
simple field theoretic interpretation, but has to be calculated as an
analytic continuation of REE, i.e. $S=\lim_{n\to 1} S^{(n)}$. 
We posit a similar $m$-particle decomposition for $S$,
\begin{equation}
	S=S_{1\PP}+S_{2\PP}+S_{3\PP}+\cdots,
\end{equation}
where $S_{m\PP}=\lim_{n\to 1} S^{(n)}_{m\PP}$. 
We calculate $S_{1\PP}$ exactly, and
show that a large class of diagrams for $S^{(n)}_{m\PP}$ vanish when the
$n \to 1$ limit is taken. For $S_{2\PP}$, we explicitly
calculate the first non-trivial diagram which involves two 2-particle
connected correlators. Note that our prescription is completely
agnostic to how the connected correlation functions are
calculated. Our formulation thus opens up the possibility of
computing EE using approximate analytic techniques,
numerically obtained correlators or even experimentally measured
ones. We believe this construction will allow large scale calculations
of EE in higher dimensions ( $d>1$) both in and out of
equilibrium.

\begin{figure*}[t!]
	\centering
	\includegraphics[trim=5cm 17.3cm 4.5cm 4cm,clip,width=1.5\columnwidth,]{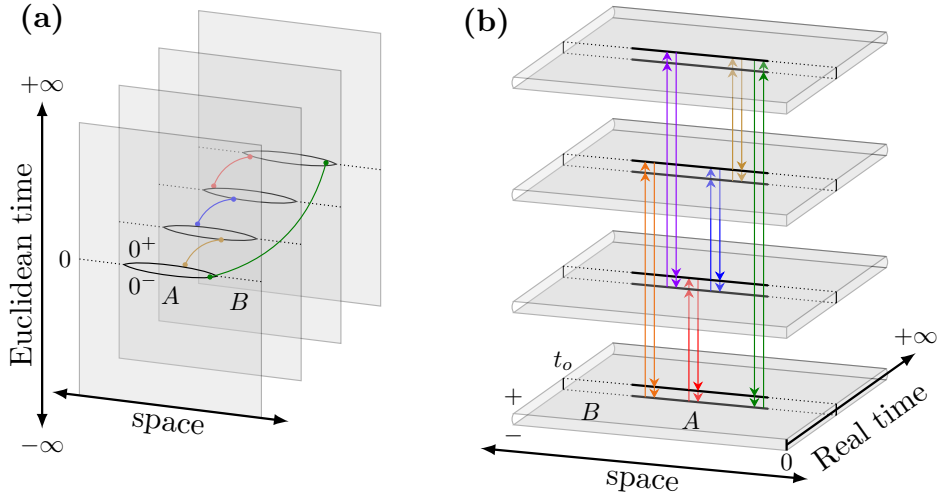}
	\caption{
		Replica based evaluation of $S^{(4)}$ : \tbf{(a)}
		Standard technique for ground states where the reduced
		density matrix is represented as a Euclidean-time path
		integral with boundary conditions in the subsystem $A$. The
		fields are matched across consecutive replica as
		shown. \textbf{(b)} Our formalism expresses each replica of
		the density matrix on a \emph{real time} Keldysh contour
		with doubled fields. Fields at each point of the subsystem
		$A$ in one replica are coupled to their counterparts in all
		other replicas. For $S^{(4)}$ this implies coupling between
		sheet $1$ and $3$, and between sheet $2$ and $4$, in
		addition to the coupling between consecutive ones. The
		coupling operates only at the time of observation $t_o$, and
		is ``directional'' in Replica space, acting as a ``Replica
		current''. Partial tracing is achieved by setting the
		coupling to zero in the complementary subsystem $B$. 
		No field matchings are required in this new formalism.
	}
	\label{fig:sourcematchings}
\end{figure*}
The remainder of the text is organised as follows. In Section \ref{EE_Free} we establish that the $n$\txup{th} order REE for a generic system of interacting fermions is equal to the Schwinger Keldysh free energy of $n$ replicas of the system coupled via inter-replica currents. We use this fact in Section \ref{WignerKeldysh} to show how $S^{(n)}$ can be decomposed into $m$ particle contributions $S^{(n)}_{m\PP}$ and provide explicit diagrammatic rules to construct the same in terms of $k$ particle correlators with $1\leq k\leq m$. Section \ref{Svn} is devoted to the analytic continuation of these contributions as $n\to1$. In Section \ref{alternate} we provide an alternate diagrammatic prescription in terms of Green's functions of the non-interacting replicas in presence of the quadratic inter-replica current terms. These correlators depend on replica indices and do not admit the usual physical interpretations as correlators in standard Keldysh filed theory. Lastly we conclude in Section~\ref{Conclusions} with a summary of our findings.

\section{ Entanglement Entropy as a Free Energy in presence of
  Replica Currents}\label{EE_Free}

Consider a system of Fermions which is made up of two mutually
exclusive spatial subsystems $A$ and $B$. For our purpose, we consider
a lattice with $V$ sites, where the subsystem $A$ has $V_A$ sites. We
will explicitly work with discrete lattice systems, and take
appropriate continuum
limit at the end. The Hilbert space of the
system is a product of the Hilbert space of degrees of freedom lying
in A and those lying in B, i.e. ${\cal H}_{AB}= {\cal H}_A \otimes {\cal
  H}_B$. On tracing over degrees of freedom in $B$, a quantum state
of the full system, $\ket{\psi(t_o)}$ gives rise to a reduced density matrix $\ro_A(t_o)$ over the
subsystem $A$. Here $t_o$ is the time of observation. The $n^{\tx{th}}$ order R\'enyi entanglement entropy (REE) of this state for the
partition between $A$ and $B$ is then given by
\begin{equation}
	S^{(n)}(t_o)=-\frac{1}{n-1} \ln\Tr[\left(\ro_A(t_o)\right)^n],
\end{equation}
while the von-Neumann entanglement entropy (referred to as EE in the following) is obtained as the analytic
continuation
\begin{equation}
	S(t_o) =\lim_{n\to1}S^{(n)}(t_o) = - \Tr[\ro_A(t_o) \ln \ro_A(t_o)].
\end{equation}
In a steady
state or in equilibrium, the RDM and hence the entropies are independent of $t_o$.
 In the standard field theoretic prescription for calculation of
 $S^{(n)}$, one considers $n$ copies(replicas) of the system. The time
 evolution of each of these replicas can be written as a functional
 integral over fields, giving rise to $n$ fields for each space-time
 point. The multiplication of RDMs translates
 to imposing the boundary condition that the fields within
 subsystem $A$ in different replicas have to be matched at the time of measurement. 
 A schematic representation of this prescription is shown in Fig.~\ref{fig:sourcematchings}(a). 
 This naturally leads to integrals over constrained field
 configurations in the replica space~\cite{Calabrese_2009}. Here, we propose a modified
 scheme (schematically shown in Fig.~\ref{fig:sourcematchings}(b)) to calculate REE of fermions which uses
 Schwinger Keldysh field theory for $n$ replicas~\cite{Chakraborty.Sensarma_PRL21_NonequilibriumDynamicsRenyi,Chakraborty.Sensarma_PRA21_RenyiEntropyInteracting,moitra_entanglement_2020,Haldar.Banerjee_PRR20_EnyiEntanglementEntropy,moitra_EEbosons}. 
 The key advantage is
 that we can trade the boundary conditions in favour of ``currents''
 flowing between replicas in the subsystem A. This allows us to
 connect REE with standard correlators of the
 original non-replicated theory.        

To develop  our formalism, it is useful to define the Wigner
characteristic $\chi$ of a density matrix, as the expectation value of the fermionic displacement
operator~\cite{Cahill.Glauber_PR69_DensityOperatorsQuasiprobability,Cahill.Glauber_PRA99_DensityOperatorsFermions,moitra_entanglement_2020},
\begin{equation}\label{eq:chi}
	\chi(\ve{\zbar},\ve{\ze};t_o)=\Tr[\ro(t_o)\e^{\sum_{i}\cd_i\ze_i-\zbar_i\cc_i}],
\end{equation}
where $\ro(t_o)$ is the density matrix of the full system. 
The Wigner characteristic has the useful property that $\chi$ for the RDM is given by
\begin{equation}\label{eq:chiA}
	\chi_A(\ve{\zbar},\ve{\ze};t_o)=\Tr[\ro(t_o)\e^{\sum_{i\in A}\cd_i\ze_i-\zbar_i\cc_i}],
\end{equation}
i.e. one simply needs to restrict the support of the displacement
operator to ${\cal H}_A$~\cite{Chakraborty.Sensarma_PRL21_NonequilibriumDynamicsRenyi}. From now onwards we will
use $\ve{\ze}$ and $\ve{\et}$ to indicate vectors of Grassmann variables with support only in $A$. Note that $\chi_A$ is a Grassmann valued
function of the Grassmann fields $\ze_i,\zbar_i$ and does
not have an immediate physical interpretation. Its usefulness lies in
the fact that expectations values of operators with support in $A$ can be written as
integrals over $\chi$,
\begin{widetext}
\begin{equation}
  \ev*{\opr{O}}[t_o] \equiv \Tr[\ro_A(t_o)\opr{O}]%
  =2^{-V_A}\int \mesr{\ve{\zbar},\ve{\zeta}}%
			\mesr{\ve{\bar{\eta}},\ve{\eta}}
            \,\chi_A\left(\sqrt{2}\ve{\bar{\eta}},\sqrt{2}\ve{\eta};\,t_o\right)
            \chi_{O}
            \left(\sqrt{2}\ve{\zbar},\sqrt{2}\ve{\zeta}\right) e^{\ve{\zbar}\cdot\ve{\eta}-\ve{\ebar}\cdot\ve{\zeta}},
  \end{equation}
where $\chi_O \left(\ve{\zbar},\ve{\zeta} \right)= \Tr[
\opr{O}~\e^{\sum_{i\in A}\cd_i\ze_i-\zbar_i\cc_i}]$ is the Weyl symbol of the
operator $\opr{O}$ and the dot product $\ve{\zbar}\cdot\ve{\eta}=\sum_{i \in
  A}\zbar_i\eta_i$. It is then easy to see (substituting $\ro_A$ for
$\opr{O}$) that the $2^\tx{nd}$ R\'{e}nyi entropy is given by
\begin{equation}
  e^{-S^{(2)}(t_o)} = 2^{-V_A}\int \mesr{\ve{\zbar},\ve{\zeta}}%
			\mesr{\ve{\bar{\eta}},\ve{\eta}}
                        \chi_A\left(\sqrt{2}\ve{\bar{\eta}},\sqrt{2}\ve{\eta};\,t_o\right)
                        \chi_{A}
                        \left(\sqrt{2}\ve{\zbar},\sqrt{2}\ve{\zeta};\,t_o\right) e^{\ve{\zbar}\cdot\ve{\eta}-\ve{\bar{\eta}}\cdot\ve{\zeta}}.
  \end{equation}
Using properties of displacement operators and fermionic coherent
states, 
Ref. \cite{moitra_entanglement_2020} showed that the $n$\txup{th} order REE corresponding to
can be written as (suppressing the time index),
%
\begin{equation}
\begin{split}
\e^{-(n-1)S^{(n)}}%
= 2^{-{(n-1)}V_A}%
\int  \prod_{\alpha=1}^{n-1}%
\mesr{\ve{\zbar}^{(\alpha)},\ve{\zeta}^{(\alpha)}}%
			\mesr{\ve{\bar{\eta}}^{(\alpha)},\ve{\eta}^{(\alpha)}}%
			~\prod_{\alpha=1}^{n-1}
			\chi_A\left(\sqrt{2}\ve{\bar{\eta}}^{(\alpha)},\sqrt{2}\ve{\et}^{(\alpha)}\right)%
			~\chi_A\left(\sqrt{2}\sum\nolimits_\alpha\ve{\zbar}^{(\alpha)},\sqrt{2}\sum\nolimits_\alpha\ve{\zeta}^{(\alpha)}\right)\\%
			\exp\left(%
				{\sum_{\alpha=1}^{n-1}\ve{\zbar}^{(\alpha)}\!\cdot\ve{\eta}^{(\alpha)}-\ve{\bar{\eta}}^{(\alpha)}\!\cdot\ve{\zeta}^{(\alpha)}%
				+\sum_{\alpha>\beta}\ve{\zbar}^{(\alpha)}\!\cdot\ve{\zeta}^{(\beta)}-\ve{\bar{\zeta}}^{(\beta)}\!\cdot\ve{\zeta}^{(\alpha)}}%
			\right).
\end{split}\label{eq:Sngeneral}
\end{equation}
\end{widetext}
Here $1\leq \alpha,\beta\leq n-1$ are replica indices. Note that while there are product of $n$ replicas,
the integration is over $2(n-1)$ pairs of Grassmann variables for each
site in $A$. 

Eq.~\eqref{eq:Sngeneral} is the starting point of our attempt to find a field
theoretic construction of EE and hence a general
relation between correlators and entanglement. For this, we work with the Schwinger Keldysh field theory of
Fermions\cite{Kamenev_11_FieldTheoryNonEquilibrium}, which describes the evolution of a many body density matrix
$\ro(t)=\opr{U}(t,0)\ro(0) \opr{U}^\dagger(t,0)$ in terms of
path/functional integrals over two
Fermionic (Grassmann) fields $\psi_{\pm}(i,t)$ at each space-time
point. The $\psi_{+}$ fields are obtained from the expansion of $\opr{U}$
(i.e. forward propagation of states) while the $\psi_{-}$ fields come from the
expansion of $\opr{U}^\dagger$ (i.e. backward propagation of states), giving
'rise to a field theory on a closed contour. The partition function $\partfn$ on this contour in presence of sources $J_\pm,\bar{J}_\pm$ is
	\begin{equation*}
	\partfn[\bar{J}_\pm,J_\pm]=\!\int\!\mesr{\pbar_{\pm},\psi_{\pm}}%
	\!\e^{\ii\actn_K[\pbar_\pm, \psi_\pm]+\ii\!\int\dd{t}\bar{J}_{+}\psi_{+}-\bar{J}_{-}\psi_{-}+\tx{h.c.}},
	\end{equation*}
where the Keldysh action $\actn_K$ determines the evolution of the
correlators in the system. Note that although we have used the same notation
$\mesr{\cdot}$ to indicate integrals over the Fermion matter fields $\psi$
as well as integrals over the arguments of the Wigner characteristic
($\zeta,\eta $), the matter fields fluctuate in space-time, while the
arguments of Wigner characteristic are evaluated at $t_o$ and hence
have only spatial variations. In this section, we
do not need specific forms of the action, and will refrain from
talking about them. One obvious advantage of using a Keldysh field
theory is that one can treat both equilibrium and non-equilibrium
situations in the same footing.

\begin{figure}[t]
	\centering
	\includegraphics[width=\columnwidth,scale=0.9]{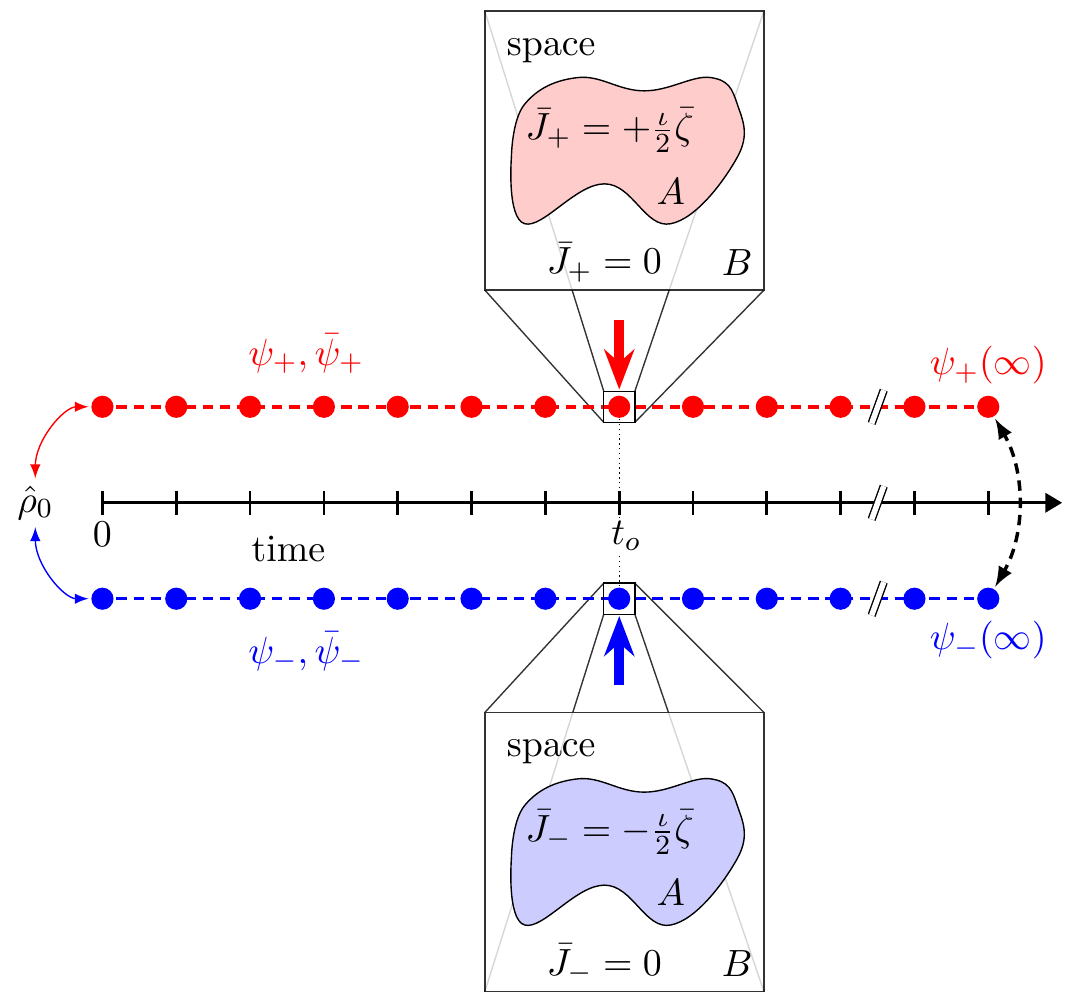} 
	\caption{Schematic representation of the Wigner
		characteristic function $\chi_A(\ve{\zbar},\ve{\ze};t_o)$
		as a Keldysh path integral with source insertions. Expanding
		the evolution operators $\opr{U}(t,0)$ and
		$\opr{U}^\dagger(t,0)$ leads to a path integral with fields
		on the ``forward''(red) and ``backward''(blue) contours
		respectively. The fields are identified at $t=\infty$ and
		are connected at $t=0$ through the matrix elements of the
		initial density matrix  $\ro_0$. The displacement operator
		is inserted symmetrically on both forward and backward
		contours at the time of observation $t_o$ and is equivalent
		to insertion of sources $J_\pm,\bar{J}_\pm$. The call-out shows the spatial
		distribution of these sources; they take nonzero values only in subsystem $A$.}
	\label{fig:keldyshcontour}
\end{figure}

In earlier works~\cite{moitra_entanglement_2020,Chakraborty.Sensarma_PRA21_RenyiEntropyInteracting,Chakraborty.Sensarma_PRL21_NonequilibriumDynamicsRenyi}, we
had shown that the Wigner characteristic of the reduced density matrix
is the Schwinger-Keldysh partition function of the system in presence
of sources which couple to fields in the
subsystem $A$ only at the time when the entropy is measured, i.e.
$\bar{J}_\pm(i,t)=\pm\,\ii\,\zbar_i/2\,\delta(t-t_o)$ if $i\in A$ and $0$ otherwise.
This is schematically shown in Fig.~\ref{fig:keldyshcontour}.  Note
that the partition function automatically traces over degrees of
freedom in $B$, the sources are inserted in $A$ to compute the Wigner characteristic.

At this point, it will be useful to shift to symmetric
($\psi_\tx{s}, \pbar_\tx{s}$) and antisymmetric ($\psi_\tx{a},\pbar_\tx{a}$) 
combinations of the $\pm$ fields defined at each spacetime point as,
\begin{align*}
	\psi_\tx{s}&=(\psi_++\psi_-)/\sqrt{2},\quad&\pbar_\tx{s}&=(\pbar_+ + \pbar_-)/\sqrt{2},\\
	\psi_\tx{a}&=(\psi_+ -\psi_-)/\sqrt{2},\quad&\pbar_\tx{a}&=(\pbar_+ - \pbar_-)/\sqrt{2}.
\end{align*} 
The symmetric source $J_\tx{s}= (J_++J_-)
/\sqrt{2}$ couples to the antisymmetric field $\pbar_\tx{a}$, while
the antisymmetric source $J_\tx{a}= (J_+-J_-)
/\sqrt{2}$ couples to the symmetric field $\bar{\psi}_\tx{s}$. For the Wigner
characteristic, the source structure works out to
\begin{equation}\label{eq:sources}
\left.
	\begin{aligned}
		J_\tx{s}(i,t)&=0,\quad&J_\tx{a}(i,t)&= -\tfrac{\ii}{\sqrt{2}}\zeta_i\delta(t-t_o),\\
		\bar{J}_\tx{s}(i,t)&=0,\quad&\bar{J}_\tx{a}(i,t)&=\tfrac{\ii}{\sqrt{2}} \zbar_i\delta(t-t_o)
	\end{aligned}
\right\rbrace\tx{ for $i \in A$,} 
\end{equation}
and $0$ otherwise. We can then write
	\begin{equation}\label{eq:chi_action}
		\begin{aligned}
			&\chi_A(\sqrt{2}\ve{\zbar},\sqrt{2}\ve{\ze}, t_o)\\%
			&=\!\!\int \mesr{\ve{\pbar}_\tx{s,a},\ve{\psi}_\tx{s,a}}%
			\e^{\ii\actn_K[\ve{\pbar}_\tx{s,a},\ve{\psi}_\tx{s,a}]%
				+\ve{\zbar}\cdot\opr{P}_A\ve{\psi}_\tx{s}(t_o)
				-\ve{\pbar}_\tx{s}(t_o)\opr{P}_A\cdot\ve{\ze}
			},
		\end{aligned}
	\end{equation}
where $\opr{P}_A$ is a projection operator on to the degrees of freedom in subsystem $A$, added to explicitly account for the sources coupling only to the degrees of freedom in $A$.
\begin{widetext}
Substituting Eq.~\eqref{eq:chi_action} into Eq.~\eqref{eq:Sngeneral}, we obtain
\begin{equation}\label{eq:Snfields}
	\e^{-(n-1)[S^{(n)}-V_A\ln2]}=%
		\int\prod_{\alpha=1}^{n}%
		\mesr{\ve{\pbar}^{(\alpha)}_\tx{s,a},\ve{\psi}^{(\alpha)}_\tx{s,a}}
		\e^{\ii\sum\nolimits_{\alpha=1}^{n}
			\actn_K\left[\ve{\pbar}^{(\alpha)}_\tx{s,a},\ve{\psi}^{(\alpha)}_\tx{s,a}\right]}%
		\e^{\ii\actn_\tx{ent}\left[
			\left\{\opr{P}_A\ve{\pbar}_\tx{s}^{(\alpha)}(t_o),\opr{P}_A\ve{\psi}_\tx{s}^{(\beta)}(t_o)\right\}
			\right]},
\end{equation}
where the \emph{entangling action}, $\actn_\tx{ent}$, is given by
	\begin{equation}
\begin{aligned}
	\e^{\ii\actn_\tx{ent}
	}&=%
	\int  \prod_{\alpha=1}^{n-1}%
	\mesr{\ve{\zbar}^{(\alpha)},\ve{\zeta}^{(\alpha)},\ve{\bar{\eta}}^{(\alpha)},\ve{\eta}^{(\alpha)}}%
	\exp\left(%
	{\sum_{\alpha=1}^{n-1}\ve{\zbar}^{(\alpha)}\!\cdot\ve{\eta}^{(\alpha)}-\ve{\bar{\eta}}^{(\alpha)}\!\cdot\ve{\zeta}^{(\alpha)}%
		+\sum_{\alpha>\beta}\ve{\zbar}^{(\alpha)}\!\cdot\ve{\zeta}^{(\beta)}-\ve{\bar{\zeta}}^{(\beta)}\!\cdot\ve{\zeta}^{(\alpha)}}%
	\right)\no\\
	&\hphantom{=\int\prod_{\alpha=1}^{n-1}}%
	\exp\left(%
	{\sum_{\alpha=1}^{n-1}%
		\ve{\ebar}^{(\alpha)}\!\cdot\opr{P}_A\ve{\psi}_\tx{s}^{(\alpha)}(t_o)%
		-\ve{\pbar}^{(\alpha)}_\tx{s}(t_o)\opr{P}_A\!\cdot\ve{\et}^{(\alpha)}
	}%
	+{\sum_{\alpha=1}^{n-1}%
		\ve{\zbar}^{(\alpha)}\!\cdot\opr{P}_A\ve{\psi}_\tx{s}^{(n)}(t_o)%
		-\ve{\pbar}^{(n)}_\tx{s}(t_o)\opr{P}_A\!\cdot\ve{\ze}^{(\alpha)}%
	}\right).%
\end{aligned}
\end{equation}
\end{widetext}
The first part of Eq.~\eqref{eq:Snfields} represents the action of $n$ independent
replicas, while the entangling action involves integrals over the source terms
in each of the replicas. Note that in absence of $\actn_\tx{ent}$, the
Keldysh partition function of the independent replicas is $1$, and
hence the entanglement would have vanished, i.e. the finite
entanglement comes solely from the effects of $\actn_\tx{ent}$. This is consistent with the
fact that $\actn_\tx{ent}$ has information about the subsystem $A$,
since the sources are restricted to $A$. While this may seem similar to the standard path
integral over $n$ replicas used in field theoretic treatment of
entanglement~\cite{casini_J.Phys.A:Math.Theor.2009_EntanglementEntropyFree,Calabrese_2009}, we note that we have $2n$ fields here. Beyond the
obvious fact that this allows us to look at non-equilibrium dynamics,
we will see that even for equilibrium systems or ground states,  using
$2n$ fields gives us certain advantages in terms of the structure of
the resultant theory.

For a non-interacting system, with Gaussian action, both
the integrals over the fields and the sources can be performed exactly
(in either order); however for a generic interacting system, it is not
possible to integrate out the matter fields exactly. Fortunately, even in
this case, the
integral over the sources is just a Gaussian integral, which can
be done analytically to get
\begin{equation}\label{eq:entanglingS}
	\ii\Sent=
	\begin{bmatrix}
		\ve{\pbar_\tx{s}}^{(1)},\ve{\pbar_\tx{s}}^{(2)},\cdots,\ve{\pbar_\tx{s}}^{(n)}
	\end{bmatrix}
	\mathbb{J}\otimes\opr{P}_A
	\begin{bmatrix}
		\ve{\psi_\tx{s}}^{(1)}\\
		\ve{\psi_\tx{s}}^{(2)}\\
		\vdots\\
		\ve{\psi_\tx{s}}^{(n)}
	\end{bmatrix},
\end{equation}
where $\mathbb{J}$ is an $n\times n$ antisymmetric matrix with all
entries below the diagonal equal to one, 
\begin{equation}\label{eq:Jmat}
	\mathbb{J}=%
	\begin{pmatrix}
		\hphantom{-}0&-1&\hphantom{-}\cdots&-1\\
		\hphantom{-}1&\hphantom{-}0&\hphantom{-}\ddots&\hphantom{-}\vdots\\
		\hphantom{-}\vdots&\hphantom{-}\ddots&\hphantom{-}\ddots&-1\\
		\hphantom{-}1&\hphantom{-}\cdots&\hphantom{-}1&\hphantom{-}0
	\end{pmatrix}_{n\times n}.
\end{equation}
Note that the integral over the sources couples the fields in different
replicas in $\actn_\tx{ent}$. It is useful to note the
following features of the entangling action:
\begin{itemize}
\item The integration of sources produces a replica coupling term
  which is quadratic in the fields and only couples fields in the
  subsystem $A$ across replicas.
 
\item The coupling is local in space and time, i.e. it connects
    fields on the same lattice sites, only at the time of measuring
    the REE. 
%
    
\item The structure of $\actn_\tx{ent}$ has the form
      of a ``current'' in the replica space, flowing between the same
      sites in the subsystem $A$ across any two distinct replicas. The entangling action does
      not couple fields in the same replica, but couples fields in all
      distinct replica pairs.
      In contrast, the standard replica trick with Euclidean field theory involves 
      fields in one replica being identified with their counterparts
      in consecutive replicas. 
      The schematic difference between the
      standard approach and our approach is shown in
      Fig.~\ref{fig:sourcematchings}(a) and (b). The use of the two-contour SK
      field theory allows the replacement of the boundary condition by
      quadratic terms.
      The boundary condition in standard replica methods requires one
      to work with actions defined on $n$-sheeted Riemann surfaces for
      generic interacting systems,
      with reduction to a single sheet possible in special cases, like in free theories ~\cite{casini_J.Phys.A:Math.Theor.2009_EntanglementEntropyFree} or in presence of conformal symmetry 
      in $1+1$D ~\cite{Calabrese_2009}.
      Our formalism on the other hand, involves ``current''-like couplings between replicas without additional boundary conditions on fields
      across replicas. This will allow us to write down a diagrammatic
      expansion for EE in terms of the local
      connected correlators of the single replica for a generic theory. We will take up this task in the next section.

\item It is important to note that the Keldysh indices
        of the fields making up the entangling action has the
        form $\sim \pbar^{(\alpha)}_\tx{s}\psi^{(\beta)}_\tx{s}$ with $\alpha\neq\beta$. 
    	Such a term involving fields from the same replica is forbidden in usual Schwinger Keldysh 
    	field theory as it would change normalization of $\ro$ and affect the causal structure of the
    	correlations.
    	Thus there is no obvious analogue of this term
        in usual single component field theories, or thermal field
        theories. The doubling of the fields in the Keldysh formalism
        is what allows us to write this term, even if we are
        considering the entanglement of a thermal state or a ground
        state. Thus, Keldysh field theory is not just a way to access
        non-equilibrium dynamics in this case, it is essential to
        writing down a space-time local description of entanglement
        entropies.

\item Finally we would like to note that the equations derived
          till now are agnostic to the explicit form of the action. For e.g.,
          they do not require conformal invariance, and can be used
          for generic interacting systems in any dimension both in and out of equilibrium.
\end{itemize}


Thus the $n$\txup{th} order REE is the Keldysh free energy of the
$n$-fold replicated system in presence of the inter-replica
``current'' between the same sites in the subsystem $A$. These
currents are active only at the time of measurement of the entropy of
the system. For a system in ground state/ thermal equilibrium / steady
state, the entropy is independent of the time of measurement, and
we can choose the time to simplify calculations. For non-equilibrium
dynamics, which tracks the time evolution of entanglement
entropy, one is interested in calculating the entropy as a function of
the measurement time $t_o$. 
We have thus provided an alternate field
theoretic picture of REE of fermionic systems. We
note that a similar construction for Bosonic systems runs into issues of
zero modes, and we will take it up in a future work~\cite{moitra_EEbosons}.

 \section{Feynman Diagrams and ``m'' particle
   Entanglement}\label{WignerKeldysh}
 
\begin{figure*}[t!] 
	\centering
	\includegraphics[width=\textwidth]{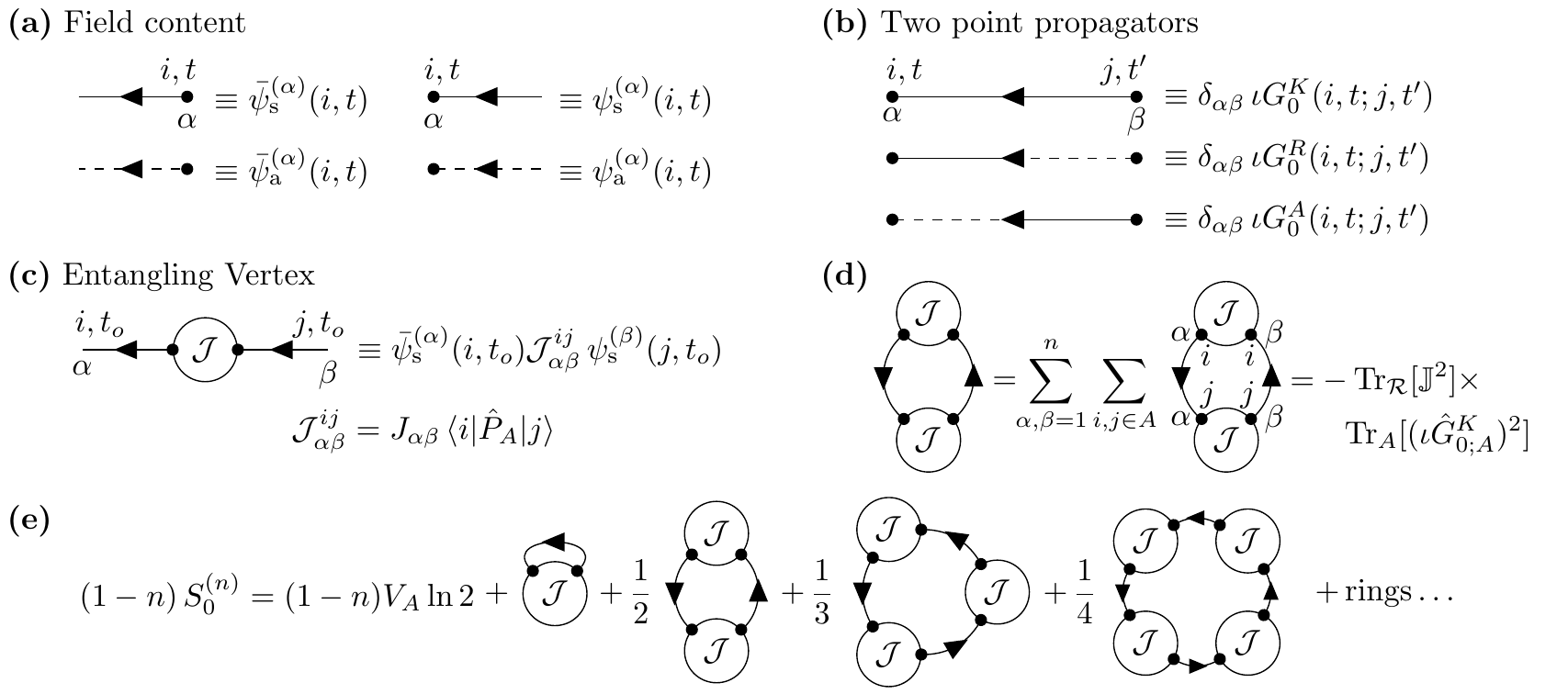} 
	\caption{
		Diagrammatic representation of \tbf{(a)} the symmetric($\psi_\tx{s}$) and antisymmetric($\psi_\tx{a}$) fields in the theory, \tbf{(b)} the two point propagators of the free theory in each replica, and \tbf{(c)} the entangling vertex (for details about vertex factors refer to main text, Eq.~\eqref{eq:entanglingS} \& Eq.~\eqref{eq:Jmat}). Lattice sites are labelled by $i,j$ and $1\leq \alpha,\beta\leq n$ are replica indices. %
		\tbf{(d)} A ring diagram with two $\cal J$ vertices. We explicitly evaluate the diagram with lattice and replica indices marked on each propagator. Since the latter are same in each replica, the resulting sum factorises into a trace over replica indices ($\cal R$), and a trace over the subsystem ($A$). 
		\tbf{(e)} R\'{e}nyi entropy in a free theory, $S^{(n)}_{0}$, is the sum of all ring diagrams. Unmarked propagators imply trace over replica and lattice indices. The term with a single $\cal J$ vertex evaluates to zero, but is still explicitly mentioned to emphasize the series structure of the ring diagrams.
	}
	\label{fig:S1p}
\end{figure*} 


In the previous section, we have shown that the $n$\txup{th} order REE 
of a system of interacting fermions is the Keldysh free
energy of $n$ replicas of the system with inter-replica currents
flowing between the sites in the subsystem $A$. Here we will try to
develop a diagrammatic expansion which will provide a prescription to
construct the REE in terms of the correlation
functions of the interacting system \emph{in a single replica sheet} 
restricted to the subsystem $A$.  More precisely, we will show that 
the $n$\txup{th} REE can be written as
 \begin{equation}\label{eq:Smp}
 	\begin{multlined}
 		S^{(n)}=%
 		S^{(n)}_{1\PP}\left[G^{(1)}\right]%
 		\,+\,S^{(n)}_{2\PP}\left[G^{(1)},G^{(2)}_c\right]\\%
 		\,+\,S^{(n)}_{3\PP}\left[G^{(1)},G^{(2)}_c,G^{(3)}_c\right]%
 		\,+\,\cdots,
 	\end{multlined}
 \end{equation}
where $G^{(1)}= -\ii\ev*{\psi_\tx{s} \pbar_\tx{s}}$ is the one-particle
(two-point) Keldysh Green's function at equal times, and 
$G^{(m)}_c=
(-\ii)^m \ev*{\psi_\tx{s} \psi_\tx{s}\cdot\cdot\,{m\rm\ times}~\pbar_\tx{s}\pbar_\tx{s}
\cdot\cdot\,{m\rm\ times}}$ 
is the equal time $m$-particle (or $2m$-point)
connected correlator of the symmetric fields at the time of
observation. 
The spatial indices of the correlators will only span
the sites in the subsystem $A$. 
	As alluded to in Eq.~\eqref{eq:Gmc.operators}, these correlators can be expressed in terms of (the connected pieces of)
	expectation values of normal ordered strings of Fermion creation and annihilation operators in the system. 
	We derive the exact form for the same in Appendix~\ref{sec:correlators}.
In the construction of Eq.~\eqref{eq:Smp}, $S^{(n)}_{m\PP}$ is independent
of $k$-particle connected correlators for $k>m$. Further, if the $m$-particle
correlator $G^{(m)}$ is factorizable, i.e. $G^{(m)}_c=0$, then
$S^{(n)}_{m\PP}$ vanishes identically. We will provide an explicit
analytic form of $S^{(n)}_{1\PP}[G^{(1)}]$ and formulate diagrammatic
rules for constructing $S^{(n)}_{m\PP}$ for general $m$. The key point
of this formulation is that it is agnostic to how the correlators are
calculated/measured and provides a recipe to stitch together the exact
correlators of the interacting theory to construct the entanglement
entropy. One can calculate the correlators in some specified
approximation scheme, but one can also substitute in this expansion 
experimentally measured correlators, or the same measured in numerical experiments like
Monte Carlo methods. This thus provides an
alternative to constructing REE from the knowledge of the exact quantum state. 
Note that reconstructing a generic many-body quantum state
requires the knowledge of an exponential number of
variables ${\cal O}(e^{V})$, while $S_{m\PP}^{(n)}$ only requires knowledge of 
upto $m$ particle connected correlators which have $\order{V_A^m}$ variables. 
Further, REE and EE are non-linear non-local functions of the RDM; there are very
few ways of computing them, whereas a large number of analytic and
numerical approximation schemes are known for calculating correlation
functions. This connection between correlators and entanglement in
Fermionic systems will allow these approximation methods to be used
in calculating REE of interacting fermions.

\begin{figure*}[t!] 
	\centering
	\includegraphics[width=\textwidth]{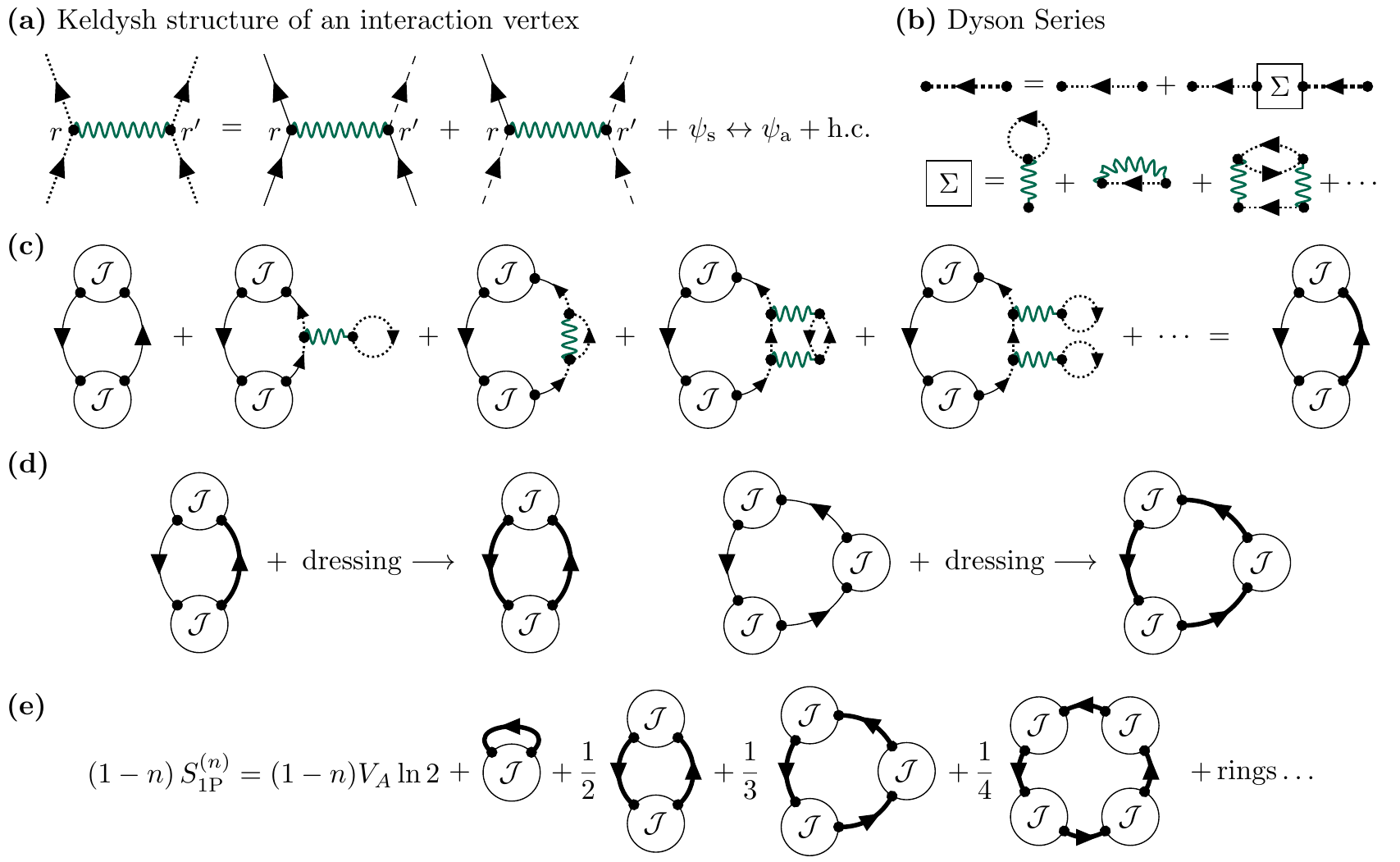} 
	\caption{
		\tbf{(a)} Interaction vertices in Keldysh field theory
		corresponding to a pair wise interaction
		$\pbar_r\pbar_{r'}U(r,r')\psi_{r'}\psi_{r}$. There are eight
		vertices in the $\psi_\mathrm{s,a}$ basis, each with
		an odd number of symmetric and antisymmetric fields,
		and a vertex factor of $\ii U(r,r')/2$ (represented as
		a green wavy line). We use fields with dotted lines as
		a placeholder for both symmetric(solid) and
		antisymmetric(dashed) fields for the sake of brevity.%
		\tbf{(b)} The presence of interactions induce self
		energy corrections $\Sigma$ which in turn dress the
		propagators. We use thick solid lines to signify the
		interacting Keldysh propagators. %
		\tbf{(c)} \& \tbf{(d)}  Interaction dressing converts the free propagators in the ring diagrams to their interacting counterparts. %
		\tbf{(e)} The one-particle contribution to $S^{(n)}$,
		$S^{(n)}_{1\PP}$ is entirely determined by the sum of
		ring diagrams with the interacting Keldysh propagators.
	}
	\label{fig:intcorrection1P}
\end{figure*}
\subsection{ One-particle correlators and $S^{(n)}_{1\PP}$}\label{sec:free1P}

We will first focus on calculating $S^{(n)}_{1\PP}$. To do this, let us first consider the 
REE of a system of free fermions, where $S^{(n)}_{m\PP}=0$ for $m>1$, i.e. $S^{(n)}=S^{(n)}_{1\PP}$. 
For the diagrammatic expansion, we denote the symmetric fermionic
fields by a solid line, and the anti-symmetric fields by dashed
lines, as shown in Fig.~\ref{fig:S1p}(a). We have omitted spin indices, but
they can be easily added to this description. Each field carries a replica index
along with usual space-time indices. We will expand the free energy around the theory
with $n$ independent identical replicas and treat $\Sent$ as an
additional coupling. In the independent replica theory, the
propagators connect fields with same replica indices. The propagators
for each replica are exactly the same as that of a single replica (
aka standard) Keldysh field theory. There are three
independent propagators, the retarded propagator $ G^R= -\ii\ev*{
	\psi_\tx{s} \pbar_\tx{a}}$,  the advanced propagator $
G^A= -\ii \ev*{\psi_\tx{a} \pbar_\tx{s}}$, and  the Keldysh propagator $
G^K= -\ii \ev*{\psi_\tx{s} \pbar_\tx{s}}$, as shown in
Fig~\ref{fig:S1p}(b). Note $\ev*{\psi_\tx{a}\pbar_\tx{a}}=0$ in the single
replica Keldysh field
theory. These definitions hold for both the interacting and
non-interacting theories. The quadratic coupling in $\Sent$
is represented by an open circle, with the ${\cal J}$ vertex given by the matrix
$\mathbb{J}\otimes\opr{P}_A$ as shown in Fig~\ref{fig:S1p}(c). 
Note that this vertex couples symmetric fields in different replicas on the same site in subsystem $A$ at the time when REE is computed.

The first point to note is that since the R\'{e}nyi entropy is a free
energy (in presence of inter-replica currents), it can be written as
a sum of all fully connected diagrams in the replica field theory with no
external legs. The second point to note is that the Keldysh free
energy of the independent replicas is $0$, so the diagrams which
contribute to REE must have one or more $\Sent$ vertices. Finally, since $\Sent$ is quadratic,
one can easily re-sum all the diagrams here. Since the diagonal
elements of $\mathbb{J}$ are $0$, there are no diagrams with a single
${\cal J}$ vertex. The first non-trivial diagram, which has two
${\cal J}$ vertices is shown in Fig.~\ref{fig:S1p}(d),
together with its explicit evaluation.  In fact, the set of ring
diagrams shown in Fig.~\ref{fig:S1p}(e) exhausts all the free energy diagrams
for the free theory. Note that only the Keldysh propagator, which
carries information about distribution functions appear in this
series. We note that the diagram with $p$ circles have a symmetry
factor of $1/p$ ( $1/p!$ from the exponential and $(p-1)!$ from
permutation of the $\cal J$ vertices). Defining $\ii\opr{G}^K_{0;A}=\opr{P}_A\ii\opr{G}^K(t_o,t_o)\opr{P}_A$ as the projection of the equal time Keldysh correlator onto the subsystem $A$, it is easy to see that
\begin{align}
	(1-n)S^{(n)}_{0}&=(1-n)V_A\ln2%
	-\sum_{p=1}^{\infty}\frac{1}{p}\Tr_{\cal R}[{\mathbb{J}}^p] \Tr_A[(\ii\opr{G}_{0;A}^{K})^{p}]\no\\%
	&=\ln(\frac{1}{2^{(n-1)V_A}}\det[\idm\otimes\id-\mathbb{J}\otimes\ii\opr{G}^{K}_{0;A}]),
\end{align}
where $\idm$ is the $n\times n$ identity matrix and $\id$ is the identitty operator on the subsystem $A$.
This determinant is known exactly (see Appendix B of Ref.~\cite{moitra_entanglement_2020}), and it reduces to a 
closed form expression for $S^{(n)}_{0}$,
\begin{equation}\label{eq:ent_nonint}
	S^{(n)}_{0}(t_o)=\frac{1}{1-n}\Tr_A\ln\left[(\opr{C}_0(t_o))^{n}+(\id-\opr{C}_0(t_o))^{n}\right],
\end{equation}
where  $C_0(i,j;t_o)\equiv\ev*{\cd_i(t_o)\cc_j(t_o)}_0$ is the non-interacting correlation matrix 
restricted to $A$, and is related to the equal-time Keldysh correlator
by $\opr{C}_0(t_o)=(\id-\ii\opr{G}^K_{0,A})/2$. Here $\cd_i$
creates a Fermion on site $i$.
Eq.~\eqref{eq:ent_nonint} matches with formulae derived earlier for non-interacting fermionic systems
~\cite{Peschel_2003,Peschel_2009,moitra_entanglement_2020}. This formula for REE of
free fermions in terms of the one-particle correlation function is
well known in the literature, and this acts as a check on our new
method to calculate entanglement.

Let us now turn our focus to interacting Fermionic systems, where, for
the sake of convenience, we assume a pairwise interaction, 
$$\Sint^{\pm}=\pm\!\int\!\!\dd{t}\sum_{r,r'}\pbar_\pm(r,t)\pbar_\pm(r',t)\,U(r,r')\psi_\pm(r',t)\psi_\pm(r,t).$$
Here $r,r'$ run over degrees of freedom in the whole system. We note
that the final formulae we derive is agnostic to the precise form of
the interaction; however assuming a representative form of
$\Sent$ helps us draw Feynman diagrams and clearly illustrate some of
the properties of the formalism. In Sec.~\ref{sec:derivation},
we outline a derivation of the same formulae without making reference
to any particular form of interaction.
For such a pairwise form the typical interaction vertices, shown in Fig~\ref{fig:intcorrection1P}(a),
consist of three symmetric and one antisymmetric fields, or vice versa. 
Note that interaction vertex couples fields with the same replica index. It is
still true that the Keldysh free energy of $n$ independent interacting
replicas is $0$, so we only need to worry about connected diagrams
with one or more open $\cal J$ vertices. 
It is instructive to think about the modifications of the ring diagrams due to 
interaction vertices. There is a set of diagrams where the interaction vertices 
couple fields on the same propagator. Some representative diagrams are shown on
Fig.~\ref{fig:intcorrection1P}(c). These are essentially self-energy corrections to
the propagators, shown in Fig.~\ref{fig:intcorrection1P}(b).  The net effect of summing up all such diagrams is to convert the
non-interacting one-particle correlator $\opr{C}_0$ to the \emph{interacting}
one particle correlator $\opr{C}$ in the expressions given in
Eq.~\eqref{eq:ent_nonint}, as seen in
Fig.~\ref{fig:intcorrection1P}(c) and
Fig.~\ref{fig:intcorrection1P}(d). These dressed diagrams, depicted in Fig.~\ref{fig:intcorrection1P}(e), exhaust $S^{(n)}_{1\PP}$ and other possible diagrams for $S^{(n)}$ depend on higher particle correlators. 
So, for interacting Fermions, we have
\begin{align}\label{eq:ent_1p_int}
	S^{(n)}_{1\PP}&=\frac{1}{1-n} \Tr_A\ln\left[\opr{C}^n+(\id-\opr{C})^n\right].
\end{align}
This is the analytic form of $S^{(n)}_{1\PP}$ in terms of the
\emph{exact interacting} one-particle correlator (two-point function).

\begin{figure*}[t!] 
	\centering
	\includegraphics[width=\textwidth]{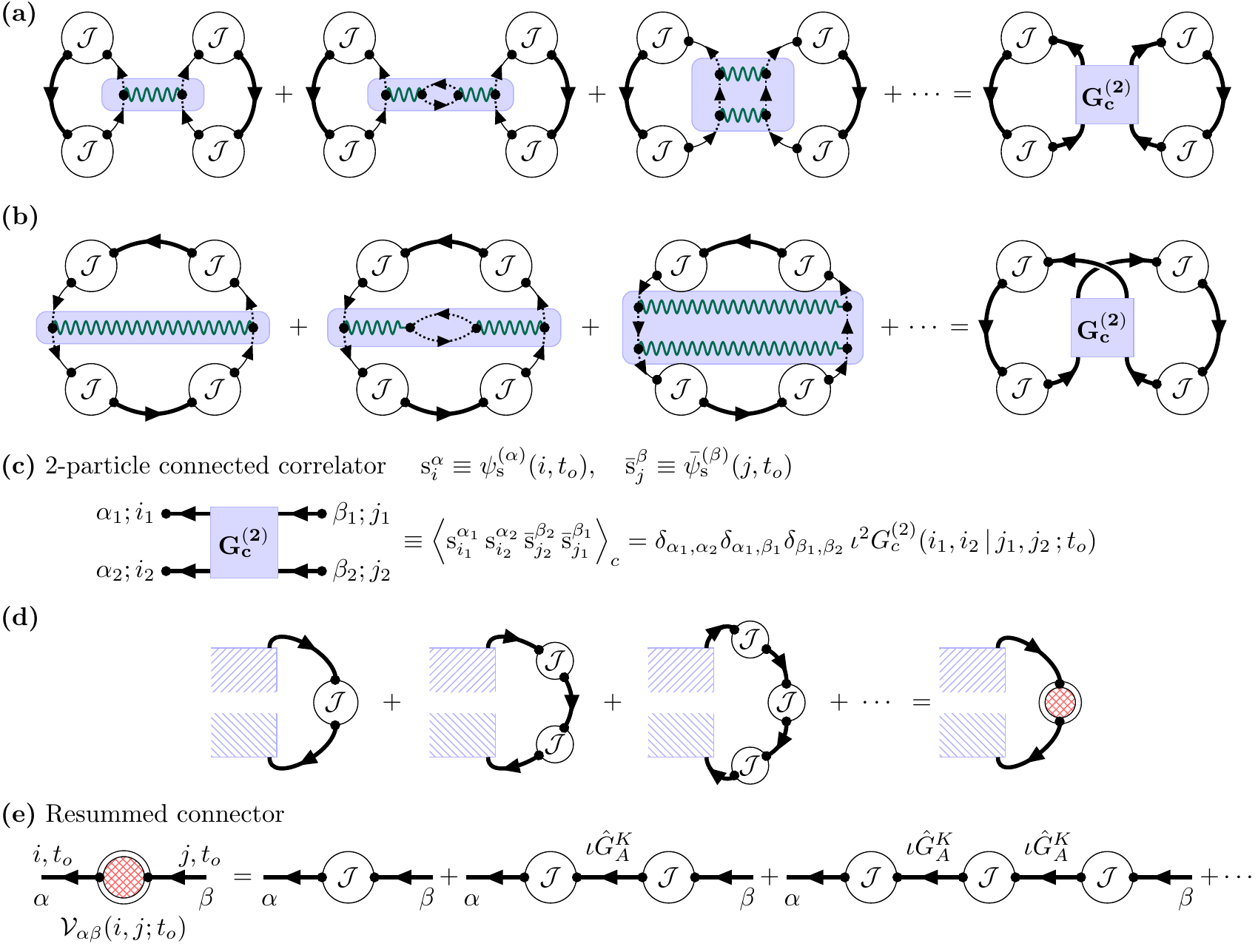} 
	\caption{Feynman diagrams for the two-particle contribution to
		REE: \tbf{(a)} Propagators in two separate rings are connected by
		interaction vertices. Dressing of this pair of propagators
		converts them to an interacting connected two-particle
		correlator $G^{(2)}_c$ (shown in \tbf{(c)}). The rings are now connected and lead
		to a valid connected diagram for $S^{(n)}_{2\PP}$. 
		\tbf{(b)} A pair of propagator within the same ring is connected by interaction
		vertices and gets dressed to $G^{(2)}_c$. 
		(a) and (b) can be viewed as diagrams where external legs of $G^{(2)}_c$ are
		connected by chains of entangling vertices ${\cal J}$ interlinked with  
		single-particle propagators. 
		\tbf{(c)} Details of the two particle
		connected correlator $G^{(2)}_c$. Note that it is diagonal in replica space.
		\tbf{(d)} Chains with different numbers of 
		entangling vertices can be resummed into a single connector ${\cal \opr{V}}$. 
		\tbf{(e)} The series for constructing ${\cal \opr{V}}$. The solid lines are interacting Keldysh
		propagators in a single replica. Detailed expressions for
		${\cal \opr{V}}$ are provided in Eq.~\eqref{eq:Vmat} and Eq.~\eqref{eq:smallv}}
	\label{fig:intcorrection_2P}
\end{figure*}

\subsection{ Two-particle correlators and $S^{(n)}_{\mathrm{2\PP}}$}

In a non-interacting system the one particle correlators define the
density matrix and hence the REE~\cite{Peschel_2009}. However this is no
longer true for interacting systems, where multi-particle connected
correlators also contribute to the REE.  To see this,
let us focus on the first diagram shown in
Fig.~\ref{fig:intcorrection_2P}(a). Here we consider two
otherwise disconnected rings, where an interaction line connects a
propagator in one ring to a propagator in the other ring. This is now
a connected diagram which contributes to $S^{(n)}$, 
but does not belong to the series of ring diagrams in $S^{(n)}_{1\PP}$.  
The set of diagrams in
Fig.~\ref{fig:intcorrection_2P} (a) shows how one can add more
complicated motifs between the pair of propagators to convert them 
to the fully interacting connected two particle
propagator (shaded region) 
\begin{align*}
	\ii^2G^{(2)}_c(i,j|k,l;t_o)&=%
	\ev{\psi_\tx{s}(i,t_o)\psi_\tx{s}(j,t_o)\pbar_\tx{s}(l,t_o)\pbar_\tx{s}(k,t_o)}_c\\
	&= 2^2\ev{ \cd_k\cd_l\cc_j\cc_i}_c.
\end{align*}
An equivalent way of getting this diagram is to first consider the
connected correlator $G^{(2)}_c$ and join its external legs through 
${\cal J}$ vertices with interacting one particle propagators in
between them. Note that each chain connecting the ends of $G^{(2)}_c$
should have at least one ${\cal J}$ vertex. Otherwise the two particle
connected correlator would effectively reduce to a single particle
propagator and reproduce diagrams in $S^{(n)}_{1P}$. It is simple to
generalize to the case of $p$ rings where pair of propagators from the
different rings are connected by interaction lines, leading to
diagrams with multiple $G^{2}_c$. A second class of diagrams
contributing to $S^{(n)}_{2P}$ can be obtained by considering pairs of
propagators in a single ring and connecting them by interaction lines,
effectively converting this pair to a $G^{(2)}_c$. A series of
connections leading to such a diagram is shown in
Fig.~\ref{fig:intcorrection_2P}(b). In both cases (a) and (b), 
the resultant diagram can be reproduced by first considering $G^{(2)}_c$ 
and connecting its legs by chains of ${\cal J}$ vertices coupled by 
one particle propagators.

\begin{figure*}[t!] 
	\centering
	\includegraphics[trim=0.3cm 0cm 1.1cm 0cm,clip,width=\textwidth]{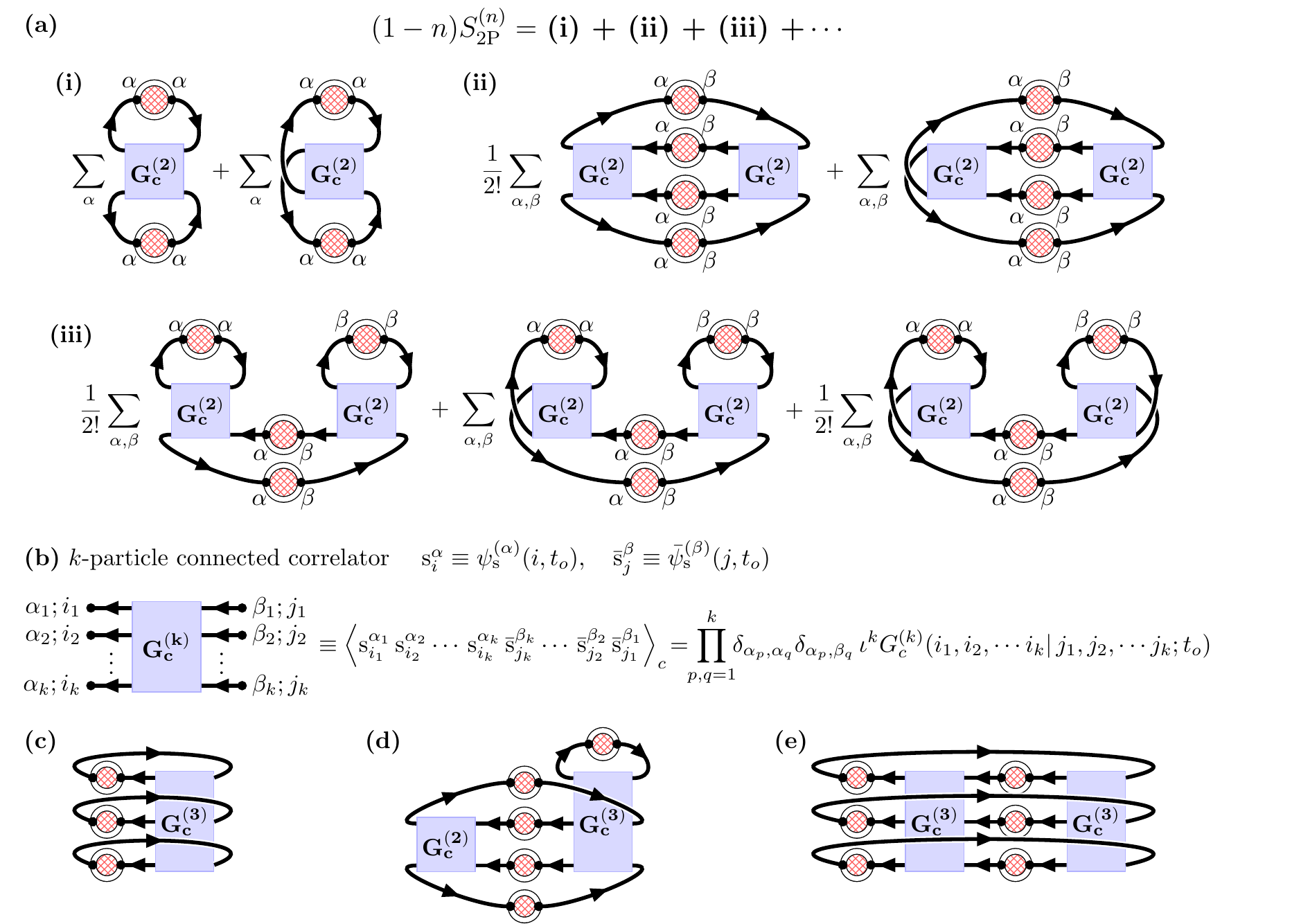} 
	\caption{\tbf{(a)} The diagrams for $S^{(n)}_{2\PP}$, which
				involve $1$ and $2$ $G^{(2)}_c$ s. 
				\tbf{(i),(ii),(iii)} The diagrams shown exhaust all topologically distinct
				possibilities up to two instances of
				$G^{(2)}_c$. $\alpha,\beta$ are replica indices, and lattice
				indices are implied to be traced over the subsystem $A$. 
				\tbf{(b)} The details of a $k$-particle connected
				correlator including replica and site indices. Note that all
				incoming and outgoing lines represent symmetric
				fields and carry the same replica index. 
				\tbf{(c)-(e)} Some representative diagrams
				contributing to $S^{(n)}_{3\PP}$. (c) involves a single
				$G^{(3)}_c$, (d) involves a $G^{(3)}_c$ and a $G^{(2)}_c$,
				while (e) involves two $G^{(3)}_c$s.
		}
	\label{fig:S2p}
\end{figure*}

At this point, it is useful to note that each chain connecting the
external legs of $G^{(2)}_c$ can have $1$, $2$, $3$, any number of ${\cal J}$
vertices. This is depicted in Fig.~\ref{fig:intcorrection_2P}(d). 
One can then use a resummed one-particle connector ${\cal \opr{V}}$ to
join the external legs of $G^{(2)}_c$ (see Sec.~\ref{sec:derivation} for details). 
The diagrammatic series corresponding to this resummation is shown in 
Fig.~\ref{fig:intcorrection_2P}(e), which evaluates to the following for $\cal \opr{V}$,
\begin{align}
	\mathcal{\opr{V}}_{\alpha\beta}&=\sum_{p=1}^{\infty}\left[{\,\mathbb{J}^p\,}\right]_{\alpha\beta}\,\left[\ii\opr{G}_A^{K}(t_o,t_o)\right]^{p-1},%
	\no\\
	&=\left[{\,\mathbb{J}\otimes\id\left(\idm\otimes\id-\mathbb{J}\otimes\ii\opr{G}^{K}_A\right)^{-1}}\right]_{\alpha\beta},
\end{align}
where $\ii\opr{G}^{K}_A=\opr{P}_A\opr{G}^{K}(t_o,t_o)\opr{P}_A$ is the interacting one particle Keldysh 
correlator at equal times, restricted to the subsystem $A$.
It is possible to analytically obtain an expression for the matrix elements of $\cal \opr{V}$, namely,
\begin{equation}\label{eq:Vmat}
	[\mathcal{\opr{V}}_{\alpha\beta}]=%
	\begin{bmatrix}
		\opr{v}_0		&-\opr{v}_{n-1}	&-\opr{v}_{n-2}		&\cdots		&-\opr{v}_{1}\\
		\opr{v}_1		&\opr{v}_0		&\ddots 			&\ddots		&\vdots\\
		\opr{v}_2		&\ddots			&\ddots				&\ddots		&-\opr{v}_{n-2}\\
		\vdots			&\ddots			&\ddots				&\opr{v}_0	&-\opr{v}_{n-1}\\
		\opr{v}_{n-1}		&\cdots			&\opr{v}_{2}		&\opr{v}_1	&\opr{v}_0
	\end{bmatrix},
\end{equation}
where
\begin{equation}\label{eq:smallv}
	\begin{aligned}
		\opr{v}_0&=\frac{1}{2}\frac{\opr{C}^{n-1}-(\id-\opr{C})^{n-1}}{\opr{C}^{n}+(\id-\opr{C})^{n}}, \\%
		\opr{v}_{k}&=\frac{1}{2}\frac{\opr{C}^{n-k-1}(\id-\opr{C})^{k-1}}{\opr{C}^{n}+(\id-\opr{C})^{n}}\tx{ for }1\leq k \leq n-1.
	\end{aligned}
\end{equation}
This resummed connector is denoted diagrammatically by two concentric open circles with the inner one filled in with a cross-hatch pattern. 
Note that while the
$\mathcal{J}(=\mathbb{J}\otimes\opr{P}_A)$ vertex had only off-diagonal components in replica space, $\mathcal{\opr{V}}$ can couple fields with same replica
indices as well. In this new language, we consider the interacting
two-particle connected correlator and join its external legs by the 
$\mathcal{\opr{V}}$ connectors to rewrite the diagrams of Fig.~\ref{fig:intcorrection_2P}(a) and (b). These are shown in Fig.~\ref{fig:S2p}(a-i).

Now, one can construct a larger class of diagrams which only
involve the fully interacting 
$G^{(2)}_c$ and $\mathcal{\opr{V}}$s. To construct these diagrams
consider $p$ different $G^{(2)}_c$ s, and connect their external legs
by $\mathcal{\opr{V}}$s, so that no external leg remains
unconnected. One can show (see Sec.~\ref{sec:derivation}) that the symmetry factor of
the diagram is simply determined by considering
the possible permutations of the $G^{(2)}_c$ blocks. All the diagrams
involving one or two $G^{(2)}_c$ blocks are shown in Fig.~\ref{fig:S2p}(a). 
The sum of all such diagrams which
depend only on the 1-particle and 2-particle correlators, 
with the number of the latter varying from
$1$ through $\infty$, forms
$S^{(n)}_{2\PP}$. 
This can be neatly summarized in the Feynman rules for diagrams in
$S^{(n)}_{2\PP}$:
\begin{itemize}
	\item Consider $p$ two particle connected correlators and join their
	external legs in all topologically distinct ways.
	\item For each connected correlator, put a factor of 
	$\ii^2 G^{(2)}_c(i,j|k,l;t_o)$, for each line joining external legs, 
	put a factor of the resummed connector ${\cal \opr{V}}_{\alpha\beta}(i,j;t_o)$.
	\item Sum over possible replica indices (noting that external legs 
	of $G^{(2)}_c$ belong to the same replica).
	\item Sum over site indices in the subsystem $A$.
	\item Multiply by $(-1)^{N_L}$, where $N_L$ is the number of
		Fermion loops.
	\item  Multiply each diagram by its symmetry factor (see Sec.~\ref{sec:derivation} for details)
\end{itemize}
Using these Feynman rules one can see that the first diagram in Fig~\ref{fig:S2p}(a-i) evaluates to
\begin{equation}
	-n\sum_{ijkl \in A} G^{(2)}_c(i,j|k,l;t_o)\,v_0(k,i;t_o)\,v_0(l,j;t_o).
\end{equation}
Similarly the first diagram (of ladder topology) shown in Fig.~\ref{fig:S2p}(a-ii) is given by
\begin{widetext}
	\begin{equation}\label{eq:ladder}
		\hphantom{-}\frac{1}{2}\sum_{\alpha\beta}%
		\sum_{\substack{ijkl\\xyzw}\in A}
		G^{(2)}_c(i,j|k,l)
		\,G^{(2)}_c(x,y|z,w)%
		\,{\cal V}_{\alpha\beta}(k,x)\,{\cal V}_{\alpha\beta}(l,y) %
		\,{\cal V}_{\beta\alpha}(z,i)\,{\cal V}_{\beta\alpha}(w,j),
	\end{equation}
	while the first diagram (of bubble topology) in Fig.~\ref{fig:S2p}(a-iii) evaluates to
	\begin{equation}
		-\frac{1}{2}\sum_{\alpha\beta}%
		\sum_{\substack{ijkl\\xyzw}\in A}
		G^{(2)}_c(i,j|k,l)
		\,G^{(2)}_c(x,y|z,w)%
		\,{\cal V}_{\alpha\alpha}(k,i)\,{\cal V}_{\beta\beta}(z,x) %
		\,{\cal V}_{\alpha\beta}(l,y)\,{\cal V}_{\beta\alpha}(w,j).
	\end{equation}
\end{widetext}
We have suppressed the time index in the above two equations for the sake of clarity. 
Summing over all such diagrams yields $S^{(n)}_{2\PP}$. 
We thus have a prescription for construction of $S^{(n)}_{2P}$ from the one-particle
and two-particle correlators.

\subsection{ $m$-particle correlators and $S^{(n)}_{m\PP}$}
Once we have established the Feynman rules for constructing the
two-particle contribution to REE, $S^{(n)}_{2\PP}$, it
is easy to extend it to the case of $S^{(n)}_{m\PP}$. Note that by
definition, diagrams for $S^{(n)}_{m\PP}$ must have at least one instance
of the $m$-particle connected correlator
\begin{equation}
	\begin{aligned}
		&\ii^m\,G^{(m)}_c(i_1,i_2,\cdots i_m|j_1,j_2,\cdots j_m; t_o) \\
		&\hspace{5pt}=
		\ev{\psi_\tx{s}(i_1,t_o)\cdots\psi_\tx{s}(i_m,t_o)\pbar_\tx{s}(j_m,t_o)\cdots\pbar_\tx{s}(j_1,t_o)}_c
		\\
		&\hspace{5pt}=(-2)^m\ev{\cd_{j_m}\cdots\cd_{j_2}\cd_{j_1}\cc_{i_1}\cc_{i_2}\cdots\cc_{i_m}}_c .
	\end{aligned}
\end{equation}
The diagrams cannot contain higher order ($k>m$)
correlators. Now, $G^{(m)}_c$ has $m$ incoming lines and $m$ outgoing
lines. Each of these outgoing lines can be (a) connected to the incoming lines
of same $G^{(m)}_c$.  Fig.~\ref{fig:S2p}(c) shows such a
diagram for $G^{(3)}_c$. (b)
connected to incoming lines of some other $G^{(k)}_c$ with
$k<m$. Fig.~\ref{fig:S2p}(d) shows such a diagram where
some of the 
outgoing lines of $G^{(3)}_c$ are connected to the incoming lines of a
$G^{(2)}_c$. 
(c) connected to incoming lines of another
$G^{(m)}_c$. Fig.~\ref{fig:S2p}(e) shows such a diagram
where two $G^{(3)}_c$ s are connected to each other. 
We draw topologically distinct connected
diagrams where all the incoming and outgoing lines are joined with $\cal\opr{V}$ vertices,
making it a free energy diagram. The
symmetry factor is determined by considering permutations of the connected
correlators in the usual way. The sum of all such diagrams give
$S^{(n)}_{m\PP}$. The Feynman rules for these diagrams are simple
extensions of those for $S^{(n)}_{2\PP}$:
\begin{itemize}
	\item Consider at least one $m$-particle connected correlator
	and any number of $k$-particle connected correlators with $k<m$. Join their
	external legs in all topologically distinct ways, so that no
	external lines remain hanging.
	\item Draw only fully connected diagrams.
	\item For each $k$-particle connected correlator, put a factor of 
	$\ii^k G^{(k)}_c(i_1,..i_k|j_1,..j_k;t_o)$, for each line joining external legs, 
	put a factor of the resummed connector ${\cal \opr{V}}_{\alpha\beta}(i,j;t_o)$.
	\item Sum over possible replica indices (noting that external legs 
	of $G^{(k)}_c$ belong to the same replica).
	\item Sum over site indices in the subsystem $A$.
	\item multiply by $(-1)^{N_L}$, where $N_L$ is the number of
	Fermion loops. 
	\item  Multiply each diagram by its symmetry factor (see Sec.~\ref{sec:derivation} for details).
\end{itemize}
This provides a general prescription to construct the $m$-particle
contribution to the R\'{e}nyi entropy $S^{(n)}_{m\PP}$. Some representative diagrams in the construction of
$S^{(n)}_{3\PP}$, which involve $G^{(2)}_c$ and $G^{(3)}_c$ are shown in
Fig.~\ref{fig:S2p}(b-d).  As an instance of employing the Feynman rules, 
the diagram in Fig.~\ref{fig:S2p}(b) is given by (time index suppressed)
\begin{equation}
	n\sum_{\substack{i,j,k\\x,y,z}\in A} \ii^3 G^{(3)}_c(i,j,k|x,y,z)\,v_0(x,i)\,v_0(y,j)\,v_0(z,k).
\end{equation}
One can now use the Feynman rules given above to construct diagrams
and convert them into integral contributions to $S^{(n)}_{m\PP}$. 
We note here for completeness that diagrams constructed following the aforementioned rules 
will contribute to $(1-n)S^{(n)}_{m\PP}$. 
We have thus provided a general prescription of constructing an
estimate of R\'{e}nyi entanglement entropy if we only have knowledge of
upto $m$-particle connected correlation functions.

\subsection{General Derivation}\label{sec:derivation}

\begin{figure*}[t!] 
	\centering
	\includegraphics[width=\textwidth]{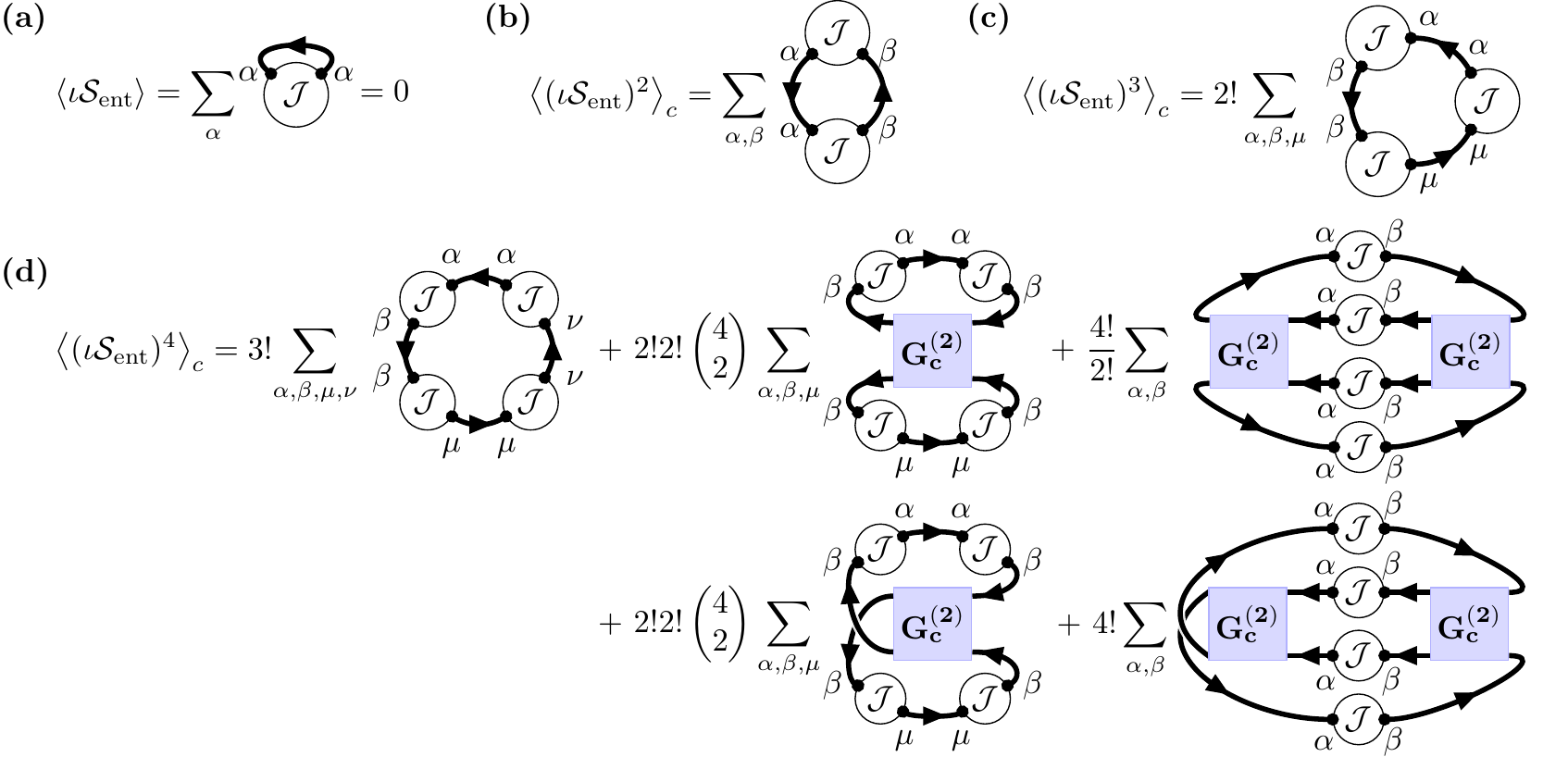} 
	\caption{ Feynman diagrams from cumulant expansion:
				\tbf{(a)-(c)} Diagrams corresponding to the first three
				cumulants of $\ii\actn_{\tx{ent}}$. Replica indices are
				explicitly marked on each propagator whereas lattice indices
				are implied to be traced over. \tbf{(d)} Diagrams
				corresponding to the fourth cumulant of
				$\ii\actn_{\tx{ent}}$. Note that this involves diagrams
				which form part of $S^{(n)}_{1\PP}$ (first diagram) as
				well as diagrams which form part of $S^{(n)}_{2\PP}$ (last 4
				diagrams). Similarly the sixth cumulant will have terms
				belonging to $S^{(n)}_{3\PP}$ as well as $S^{(n)}_{1\PP}$
				and $S^{(n)}_{2\PP}$. Note that each of $S^{(n)}_{m\PP}$ has
				contribution from an infinite set of cumulants starting at order $2m$. The cumulants 
				simply correspond to a different grouping of all the diagrams discussed earlier.}
	\label{fig:cumulants}
\end{figure*}

In the discussion so far, we started by inspecting what effect a ``replica current'' term $\Sent$ would have on the free energy diagrams of $n$ independent free theories $\actn_0$. We then proceeded to add interactions $\Sint$ in each replica to argue for, and illustrate, various non-perturbative effects. In this section we instead choose to formally expand Eq.~\eqref{eq:Snfields} in cumulants of $\ii\actn_\tx{ent}$ and evaluate them in terms of connected correlators in the (un-replicated) interacting theory $\actn_0+\Sint$. 
This will provide a first principles derivation of the previously stated results which makes no appeal to the form or nature of $\Sint$.

We crucially use three facts mentioned earlier, but reiterated here for emphasis: 
(i) REE $S^{(n)}$ is equivalent to a Keldysh free energy of $n$ independent 
replicas in presence of the entangling action $\Sent$ 
(ii) Keldysh partition functions are inherently normalised to unity in 
the absence of sources which implies that all diagrams must include at 
least one instance of $\Sent$ 
(iii) All correlators which occur in the diagrammatic expansion are ``standard'' Schwinger-Keldysh correlators of a single replica; i.e. these are the correlation functions one finds in standard textbooks \cite{Kamenev_11_FieldTheoryNonEquilibrium}. 
Owing to the structure of $\Sent$, only equal time correlators of the symmetric fields, restricted to the subsystem $A$, will make an appearance. With this in mind, $\Tr[\ro_A^n]$ can be viewed as an expectation value of $\e^{\ii\actn_\tx{ent}}$ in $n$ independent copies of the interacting theory, 
\begin{equation*}
	\e^{-(n-1)(S^{(n)}-V_A\ln2)}=\ev{\e^{\ii{\cal S}_\tx{ent}}}_n,
\end{equation*}
where the subscript on $\ev{\bullet}_n$ denotes the number of independent copies of the action $\actn_K$. The REE $S^{(n)}$ can be obtained as the cumulant expansion of the above,
\begin{align}
	&(1-n)(S^{(n)}-V_A\ln2)\no\\
	&\hphantom{-}=\ev{\ii\actn_\tx{ent}} %
	+\frac{1}{2!}\ev{(\ii\actn_\tx{ent})^2}_c%
	+\frac{1}{3!}\ev{(\ii\actn_\tx{ent})^3}_c+\cdots,\label{eq:cumulantexp}
\end{align}
where $\ev{\bullet}_c$ represents the connected part of the expectation value calculated with respect to the $n$ independent replicas.
For example, in case of the second cumulant  $\ev{(\ii\actn_\tx{ent})^2}_c=\ev{(\ii\actn_\tx{ent})^2}_n-\ev{\ii\actn_\tx{ent}}^2_n$ . We will employ the diagrammatic rules set up previously to evaluate these cumulants. A $p$\txup{th} order cumulant will involve all possible topologically distinct connected diagrams made out of $p$ number of $\mathcal{J}$ vertices. The first few orders are depicted in Fig.~\ref{fig:cumulants}. Since the $\cal J$ vertex doesn't connect fields in the same replica, the first cumulant $\ev{\ii\Sent}_n$ is exactly zero. This simplifies the diagrams at higher orders. In particular, till the third cumulant, the only possible fully connected diagrams are ring diagrams as shown in Fig.~\ref{fig:cumulants}(a)-(c). Explicitly evaluating the diagram in Fig.~\ref{fig:cumulants}(b) and (c) we get,
 	\begin{align*}
 		\ev{(\ii\actn_{\tx{ent}})^2}_c&=%
 		-\Tr_{\cal R}[\mathbb{J}^2]\Tr_A[(\ii\opr{G}^K_A)^2],\\
 	\ev{(\ii\actn_{\tx{ent}})^3}_c&=-2!\Tr_{\cal R}[\mathbb{J}^3]\Tr_A[(\ii\opr{G}^K_A)^3].
	\end{align*}
where $\ii\opr{G}^K_A=\opr{P}_A\ii\opr{G}^K(t_o,t_o)\opr{P}_A$ is the equal time two-point Keldysh correlator evaluated in the interacting theory and restricted to subsystem $A$. In a similar fashion, every higher order cumulant will give rise to a ring diagram of $\mathcal{J}$ vertices, with the number of cyclic permutations at order $p$ being $(p-1)!$. Taken together with the combinatorial factor of $1/p!$ in the definition of the cumulant expansion, the weight of a ring diagram at order $p$ is $1/p$. Grouping all order ring diagrams together we have recovered the \emph{$1$-particle contribution to $S^{(n)}$}, $S^{(n)}_{1\PP}$ as depicted in Fig.~\ref{fig:intcorrection1P}(e), and consequently the analytic form of the same in terms of the interacting correlation function $C$ in Eq.~\eqref{eq:ent_1p_int}.

In higher order cumulants, connected diagrams of $\cal J$ vertices can also be constructed using multiparticle connected correlators. For example Fig.~\ref{fig:cumulants}(d) shows the diagrams with $G^{(2)}_c$ that contribute to the fourth order cumulant. In fact, the $p$\txup{th} order cumulant will involve all possible fully connected diagrams with $p$ instances of the entangling vertex $\mathcal{J}$ joined by all possible $k$-particle equal time connected Keldysh correlators $G^{(k)}_c$ with $2k\leq p$.

We now turn to classifying the diagrams based on their correlator content. Diagrams with atleast one instance of the 2 body connected correlator $G^{(2)}_c$, but no higher body correlators are clubbed together as the ``$2$-particle contribution'', $S^{(n)}_{2\PP}$. Similarly we define the ``$m$-particle contribution'' to $S^{(n)}$ as the collection of all diagrams which contain atleast one instance of $G^{(m)}_c$ but none of the higher body correlators. This formal regrouping lets us write a ``$m$-particle''-decomposition for EE as stated in Eq.~\eqref{eq:Smp}. Given the structure of $S^{(n)}_{m\PP}$, if the $m$\txup{th} correlator is factorisable, the entire collection of diagrams get decimated.

Each of these $m$-particle contributions contain a multitude of diagrams with different correlator content 
(number and type of $G^{(m)}_c$ boxes) and topologies. They also include infinitely many diagrams of the same 
topology and correlator content but with a variable number of $\cal J$ vertices on the lines between the 
$G^{(m)}_c$ boxes. Such diagrams can be clubbed together to get a resummed connector $\cal V$ on \emph{each line} since the relative weights of the diagrams turn out to be exactly one. Assume a diagram of a given structure with a total of $p$ number of $\cal J$ vertices, distributed across $q$ different lines with the number in each line being $n_1,n_2,\cdots,n_q$. The ways of distributing $p$ identical vertices into $q$ lines with the number of vertices in each line fixed, gets exactly cancelled by the $1/p!$ from the cumulant expansion and $n_j!$ permutations from the $j$\txup{th} line, 
$$
\tx{Weight}=\frac{1}{p!}\times\frac{p!}{n_1!\,n_2!\,\cdots n_q!}\times(n_1!\,n_2!\,\cdots n_q!)=1.
$$
This fact reproduces the series depicted in Fig.~\ref{fig:intcorrection_2P}(d) and the functional forms in Eq.~\eqref{eq:smallv}. This also makes it apparent that any and all symmetry factors for the diagram is determined from exchange of the correlator blocks. In summary, all diagrams are constructed following the Feynman rules laid down in the previous sections.

We have thus provided a constructive prescription for evaluating
$S^{(n)}_{m\PP}$ in a manner independent of the underlying theory of the problem. 
We note that if one is interested in the
exact answer for entanglement, one has to compute upto
$S^{(n)}_{V_A\PP}$ and the complexity of the problem is same as exact
diagonalization. However field theoretic methods are rarely good for
exact answers, they are usually geared to provide useful approximate answers
to various quantities. In this case, the decomposition of $S^{(n)}$
into $S^{(n)}_{m\PP}$ is useful when the series
can be truncated after a few terms to yield good estimates of
entanglement. This will happen if the higher order connected
correlators are parametrically small, i.e. the system is connected to
a Gaussian theory by small couplings. This can happen in a weakly
interacting Fermi liquid, where the higher order correlators occur at
higher orders in the interaction strength. It can also happen if the
Gaussian theory represents a symmetry broken mean field state, e.g. in
a large $N$ theory of a superconductor or magnets, where higher order connected correlators have
larger powers of $1/N$. In general, if one only has information about
few body correlators, one can use this decomposition to obtain
estimates of REE. The question of specific
approximation schemes for particular situations is not discussed
here. It will be taken up in future works.

\begin{figure*}[t!]
	\centering
	\includegraphics[trim=0cm 0cm 0cm 0cm,clip,width=1.5\columnwidth,]{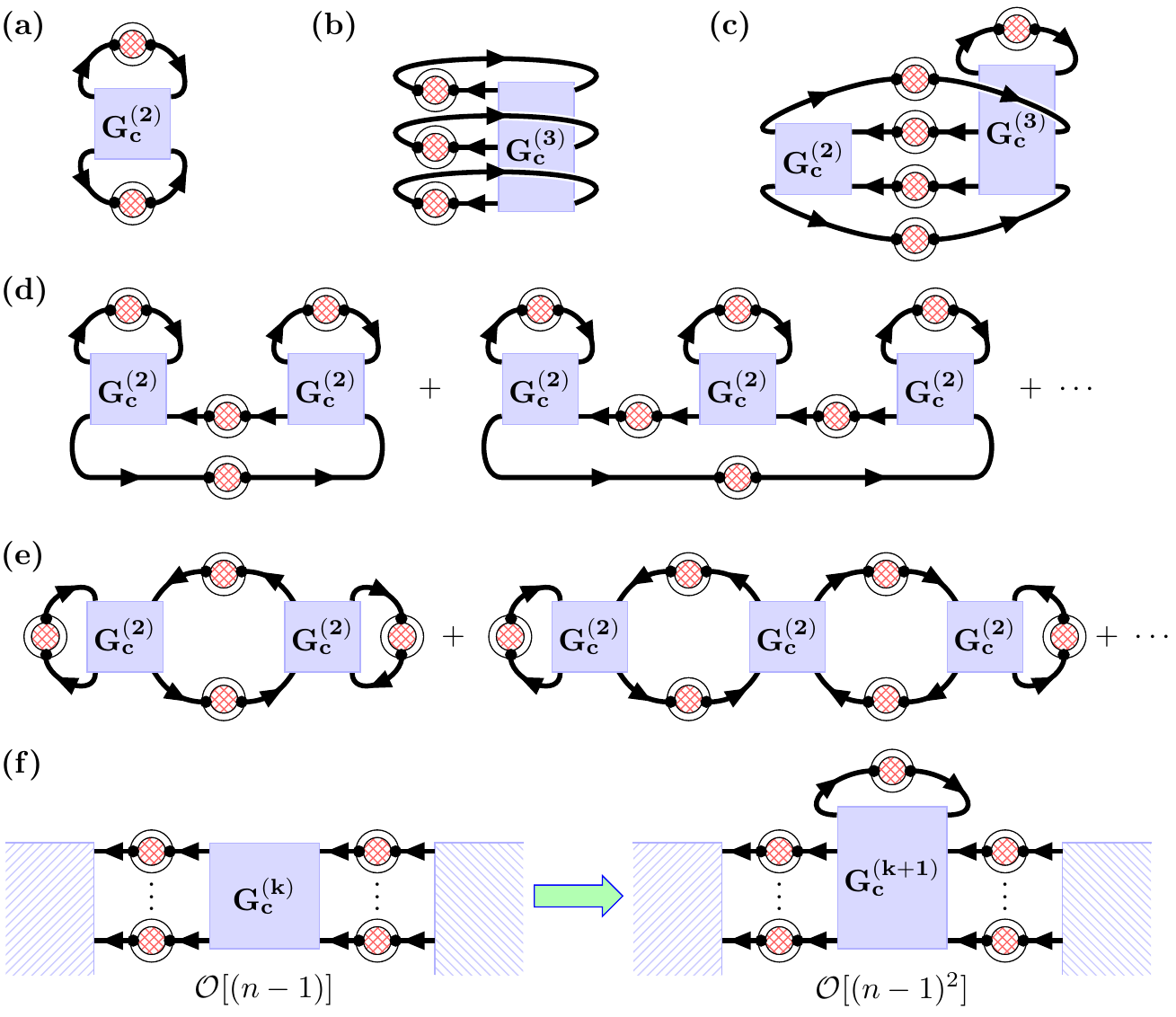}
	\caption{Examples of diagrams which do not contribute to EE, i.e. do not survive under analytic continuation $n\to 1$. 
			\tbf{(a)-(b)} Diagrams with a single multi-particle correlator with external legs connected amongst themselves through $\opr{v}_0$. (a) scales like $\sim(n-1)^2$ whereas (b) scales as $\sim(n-1)^3$.
			\tbf{(c)} Diagram with mixed correlators scaling as $\sim(n-1)^2$ despite having only one instance of $\opr{v}_0$ .
			\tbf{(d)-(e)} Series of diagrams with recurrent motifs which get decimated in the $n\to 1$ limit. In (d) each successive diagram with $p$ motifis scales as atleast $\order{(n-1)^p}$, whereas in (e) each successive diagram scales as atleast $\order{(n-1)^2}$. Note that the first diagram in both series are topologically equivalent. It is included twice to emphasize the series structure.
			\tbf{(f)} A given diagram can be modified to generate new diagrams by replacing a $k$-particle connected correlator by a $(k+1)$-particle correlator with a contracted pair of legs. The extended diagram preserves the replica structure of connectors as in the original but scales with one higher power of $n-1$ in the $n\to 1$ limit. Such extensions are ``irrelevant'' for the analytic continuation, in the sense that if the original diagram survived the $n\to 1$ limit, the extended diagram would not. The diagram in (c) is such an irrelevant extension of a diagram which has a nonzero analytic continuation, and hence has a zero contribution itself. 
		}
	\label{fig:analyticcontinuation}
\end{figure*}

\section{Analytic continuation and von-Neumann Entropy}\label{Svn}

In the previous section we have shown a way to construct the $n$\txup{th}
order R\'{e}nyi entropy in terms of the multi-particle correlators of an
interacting Fermionic system. In this section we will discuss how this
can be used to obtain a construction of the von Neumann entropy (EE) $S=
\lim_{n\to 1} S^{(n)}$. The diagrammatic construction discussed so far is given for $(1-n)S^{(n)}$. Hence, only diagrams which scale as $\sim(n-1)$ will contribute to $S$ in this expansion; terms $\sim\order{(n-1)^2}$ will not survive the analytic continuation.

Let us first consider the one-particle
contribution $S_{1\PP}$. Taking the analytic continuation of
Eq.~\eqref{eq:ent_1p_int}, one can easily show that
\begin{equation}
  S_{1\PP}=-\Tr[\opr{C}\ln\opr{C}+(\id-\opr{C})\ln(\id-\opr{C})].
\end{equation}
It is hard to formulate such a general expression for analytic continuation
of $S^{(n)}_{m\PP}$. This is primarily because of the fact that elements
of the matrix $\mathcal{\opr{V}}_{\alpha\beta}$ have explicit $n$ dependence as
well as dependence on $\alpha-\beta$. Thus the $n$ dependence of different
diagrams, which involve summing over replica indices, have to be
calculated individually for each diagram and a priori cannot be captured by a
general formula. However, a large class of diagrams vanish when the
analytic continuation is taken and the diagrammatic expansion for $S$
has many diagrams less than that for $S^{(n)}$. To see this, note that
in the limit $n \to 1$,
\begin{equation}\label{eq:v0scaling}
  \opr{v}_0 \sim \frac{n-1}{2} \ln
  \left[\frac{\opr{C}}{1-\opr{C}}\right] + \order{(n-1)^2}.
  \end{equation}
This has the immediate consequence that any term involving more than one instance of
$\opr{v}_0$ must scale at least as $\sim (n-1)^2$ and hence have vanishing contribution 
in the $n\to 1$ limit. These $\opr{v}_0$ connectors may connect the external legs on the same correlator, or the legs from different correlators with the same replica index.
As an example, consider diagrams like those in \bcol{Fig.}~\ref{fig:analyticcontinuation}(a) \& (b), where the external legs of a single multi-particle correlator are joined by the resummed connector $\mathcal{\opr{V}}$. Such diagrams necessarily contain more than one factor of $\opr{v}_0$ and are decimated under analytic continuation. 
Similarly one can show that entire series of diagrams vanish under analytic continuation due to this criterion. We depict two examples of such series in \bcol{Fig.}~\ref{fig:analyticcontinuation}(d) \& (e) which have the topology of Harteee corrections and RPA diagrams of many body theory respectively.

The leading scaling of $\opr{v}_0$ in the $n\to1$ limit can also be used to constrain possible diagrams which have non-zero contribution to EE. 
Given a legitimate diagram with a $k$-particle correlator, we can generate another equally legitimate diagram from it by replacing the chosen correlator with a $(k+1)$-particle one with a pair of external legs self-contracted through $\opr{v}_0$, as shown in Fig.~\ref{fig:analyticcontinuation}(f). This procedure keeps the replica structure of the connectors in the original diagram intact. However the presence of the extra $\opr{v}_0$ in the extended diagram increases the leading $(n-1)$ scaling by one power as compared to the original.
If the original diagram had a non-zero contribution to EE, the diagram born from such an ``irrelevant'' extension will not contribute to EE. As an example, the diagram in Fig.~\ref{fig:analyticcontinuation}(c) is an irrelevant extension of the diagram in Fig.~\ref{fig:S2p}(a-ii) and doesn't contribute in the $n\to1$ limit for EE.

We now turn to focus on diagrams which do survive in the $n\to 1$ limit.
As an example, consider the analytic continuation of the diagrams for
$S^{(n)}_{2\PP}$ shown in Fig.~\ref{fig:S2p}(a). The diagrams of Fig.~\ref{fig:S2p}(a-i) and (a-iii) vanish in the $n \to 1$ limit, while the first
non-trivial diagram for $S_{2\PP}$ is given by the analytic continuation of the
diagram in Fig~\ref{fig:S2p}(a-ii). 
For the evaluation of this diagram, it is more convenient to work in the eigenbasis of the interacting 
correlation matrix $C(i,j)$ restricted to the subsystem $A$, 
defined as $\opr{C}\ket{c_m}=c_m\ket{c_m}$ where $c_m$ and $\ket{c_m}$ 
are respectively the eigenvalues and eigenvectors of $\opr{C}$.
For $n \to 1$, this diagram gives (see Appendix~\ref{sec:appendix} for details)
\begin{equation}\label{eq:svn_twog2}
-\frac{1}{32}\sum_{c_1,c_2,c_3,c_4}\frac{\ln(x)-\ln(y)}{x-y}\left|\mel**{c_1,c_2}{\opr{G}^{(2)}_c}{c_3,c_4}\right|^2,
\end{equation}
where $x=c_1c_2(1-c_3)(1-c_4)$ and $y=(1-c_1)(1-c_2)c_3c_4$, and $\ket{c_1,c_2}=\ket{c_1}\otimes\ket{c_2}$.
The arguments presented in Appendix \ref{sec:appendix} can be readily adapted to analytically continue other diagrams with the same replica connection structure, such as the diagram in Fig.~\ref{fig:S2p}(e).

So far we have assumed that the expression in Eq.~\eqref{eq:svn_twog2} is finite. Indeed the $(\ln x-\ln y)/(x-y)$ factor in the summand of Eq.~\eqref{eq:svn_twog2} is finite for $x=y$ but diverges when either of $x,y\to0$. It is a priori unclear if the matrix elements of $G^{(2)}_c$ are sufficiently small in this regime to result in a finite answer for this diagram. In case they do not converge, the program of diagram-by-diagram analytic continuation falls under suspicion and some appropriate subset of diagrams in $S^{(n)}_{m\PP}$ might have to be resummed first and then analytically continued. Such considerations and its implications for entanglement are left as topics of future work.

\begin{figure*}[t!] 
	\centering
	\includegraphics[trim=0cm 0.6cm 0.6cm 0cm,clip,width=\textwidth]{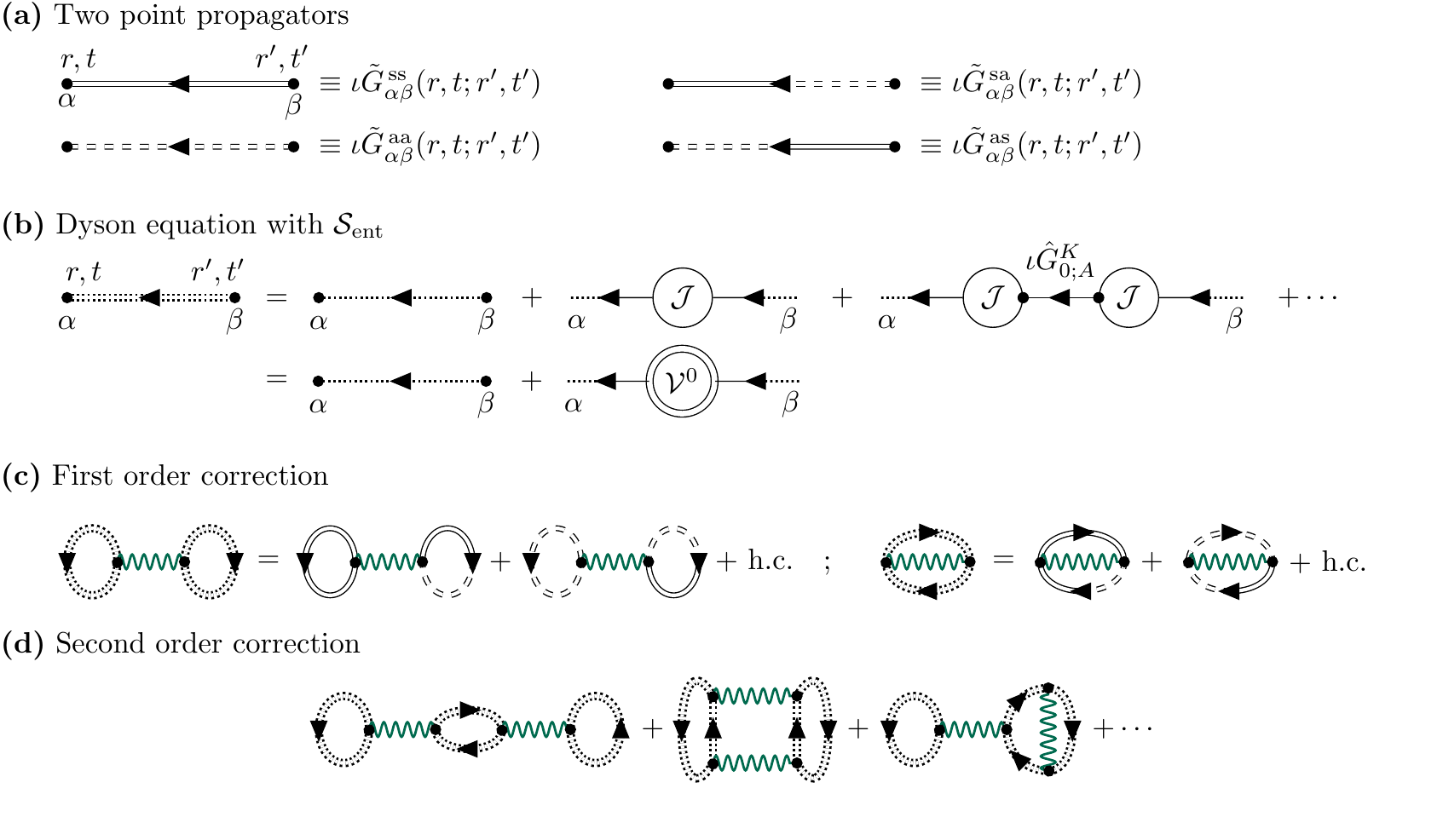} 
	\caption{
			Feynman diagrams for R\'{e}nyi entanglement entropy using
			propagators in the replicated theory. \tbf{(a)}:
			Single particle propagators $\tilde{G}$ in the non-interacting theory with the
			replica coupling term $\Sent$ included. We use double
			lines for these propagators and single lines for
			propagators in a single replica ($G$). $\alpha$ $\beta$ are the
			replica indices. Note that $\tilde{G}^{\rm\, aa}$ is finite
			in this theory, in contrast to a single replica
			Keldysh theory. %
			\tbf{(b)}: The Dyson series for the $\tilde{G}$ in
			terms of $G$ and the entanglement vertex ${\cal
				J}$. The structure of the Dyson equation is similar
			for all Keldysh components. Hence a single diagram
			with dotted lines is used to show this. The specific
			Keldysh components can easily be obtained from this by
			replacing the dotted external lines with a solid
			(symmetric) or a dashed (antisymmetric) line. 
			Note that for $\tilde{G}^{\rm\, aa}$, the first term $\ev*{\psi_\tx{a}\pbar_\tx{a}}_0=0$. %
			\tbf{(c)}: Diagrams showing the first order correction
			due to interaction vertices. The diagrams are drawn
			with double dotted lines to show their structural
			similarity with standard free energy diagrams in
			equilibrium field theory (Direct and Exchange terms).
			However, the propagators
			here carry Keldysh  indices and only particular
			combinations are allowed by interaction vertices. This
			is explicitly shown in (c) for the first order
			diagrams. Note that lines emanating
			from an interaction vertex must have the same replica
			index. %
			\tbf{(d)}: Feynman diagrams corresponding to second
			order corrections in the interaction strength. Only
			the structure of diagrams is shown with dotted lines
			for the sake of brevity.
	}
	\label{fig:dblprop}
\end{figure*}
\section{Beyond Standard Keldysh Field Theory}\label{alternate}

In section~\ref{sec:free1P}, we have shown that the $n$\txup{th} R\'{e}nyi entropy of a
system of interacting Fermions is the Keldysh free energy of n
replicas governed by an action, $\actn_0+\actn_\tx{int}+\Sent$, where $\actn_0$ is
the non-interacting (quadratic) action for decoupled replicas,
$\actn_\tx{int}$ is the interacting part of the action for decoupled replicas
and $\Sent$ is the quadratic action which couples the different
replicas and generates entanglement. In section~\ref{sec:derivation},
we expanded the free energy around $\actn_0+\actn_\tx{int}$ and connected entanglement
entropy with interacting connected correlators in the usual single replica
theory. However, one can also think of solving $\actn_0+\Sent$ exactly
since this is still a quadratic action, and use the resulting
propagators to expand in $\actn_\tx{int}$. Note that this is a regrouping of
the terms worked out earlier. In this section we will work out this
expansion and resultant diagrammatics.

The earlier expansion around $\actn_0+\actn_\tx{int}$ had the advantage that it used
propagators and correlators which can be related to observable
correlations in the system. The one-particle propagators, for example,
can be grouped into retarded, advanced or Keldysh propagators with
well known properties and relations to measured quantities. The
current expansion will work with propagators which are matrices in
replica space and cannot be immediately related to observables. They
will not follow the clear demarcation into retarded, advanced or
Keldysh propagators, at least they will not inherit their well known
properties. However, there are two advantages to the current
expansion: (i) the organization of diagrams is simpler, since the
replica space propagators already incorporate resummations due to
$\Sent$ and (ii) this expansion will tie up vertex functions rather
than correlation functions. This can be of relevance if one is interested in
calculating ``effective entanglement actions'' under various
circumstances with the ``effective entanglement action'' playing a similar role as that
of modular Hamiltonians~\cite{Dalmonte.Vermersch_AP22_EntanglementHamiltoniansField} used to study entanglement entropy. Since this maintains the language of an effective
action, this is also the natural language to introduce techniques
like auxiliary fields, saddle points etc. This is also the natural
language to think about renormalization group in this context.

Let us first focus on $\actn_0+\Sent$. In this case both the fields and
the propagators carry space-time, Keldysh and replica indices (one can
add spin or other quantum numbers as well). Using $\Sent$ as a self
energy correction, one can solve the Dyson series exactly (see
Fig.~\ref{fig:dblprop}(b)), and define the propagators
\begin{widetext}
	\begin{equation}
		\begin{aligned}
			\tilde{G}^{\rm\,sa}_{\alpha\beta}(r,t;r',t') &=%
			\delta_{\alpha\beta} G^R_0(r,t;r',t')%
			+\ii\sum_{i,j\in A}G^K_0(r,t;i,t_o)\,{\cal V}^0_{\alpha\beta}(i,j)\, G^R_0(j,t_o;r',t'),\\
			\tilde{G}^{\rm\, as}_{\alpha\beta}(r,t;r',t') &= %
			\delta_{\alpha\beta}G^A_0(r,t;r',t')%
			+\ii\,\sum_{i,j\in A}G^A_0(r,t;i,t_o)\,{\cal V}^0_{\alpha\beta}(i,j)\, G^K_0(j,t_o;r',t'),\\
			\tilde{G}^{\rm\, ss}_{\alpha\beta}(r,t;r',t')&=%
			\delta_{\alpha\beta}G^K_0(r,t;r',t')%
			+\ii\,\sum_{i,j \in A}G^K_0(r,t;i,t_o)\,{\cal V}^0_{\alpha\beta}(i,j)\, G^K_0(j,t_o;r',t'),\\
			\tilde{G}^{\rm\, aa}_{\alpha\beta}(r,t;r',t') &=%
			\ii\,\sum_{i,j \in A}G^A_0(r,t;i,t_o)\,{\cal V}^0_{\alpha\beta}(i,j)\, G^R_0(j,t_o;r',t'),
		\end{aligned}
	\end{equation}
\end{widetext}
where $G^R_0$, $G^A_0$ and $G^K_0$ are the retarded, advanced, and Keldysh
propagators of the single replica Keldysh field theory without interactions.
$\mathcal{\opr{V}}^{0}$ has the same structure as in Eq.~\eqref{eq:Vmat} but with the non-interacting correlation matrix determining the blocks in Eq.~\eqref{eq:smallv}. 
We denote this difference by using concentric empty circles to represent $\cal \opr{V}_0$ 
in contrast to the cross-hatched concentric circles used for $\cal \opr{V}$.
These propagators and the Dyson series for them is shown in
Fig.~\ref{fig:dblprop}(a) and (b). Although these propagators are not directly related to observables, they can be
constructed out of the standard one-particle propagators. We note that
in this case $\tilde{G}^{\rm\, sa}$ is not a retarded propagator, nor is
$\tilde{G}^{\rm\, as}$ an advanced propagator, although the relation
$\left(\tilde{G}^{\rm\, as}\right)^\dagger =\tilde{G}^{\rm\,sa}$ still
holds. Further, unlike the standard Keldysh theory, $\tilde{G}^{\rm\,aa}$
is non-zero due to the presence of $\Sent$ in the action. 
It is easy to show that $G^{\rm\,ss}$ and $G^{\rm\,aa}$ are
both anti-Hermitian in this replicated theory. These propagators are
represented by a double line in the diagrams.

One can now work out the diagrammatic expansion of the free energy in
terms of the original interaction vertices in $\Sint$ and the new
propagators in the usual way:(a) draw all topologically distinct
connected diagrams. (b) For each interaction vertex, put $\ii U(r,r')/2$,
where $U(r,r')$ is the  matrix
element of the interaction, for each propagator put a factor of
$\ii\tilde{G}$. (c) multiply by symmetry factor and $(-1)^{n_L}$, where
$n_L$ is the number of Fermion loops (d) Sum over all internal indices (over all space and time, \emph{not} only in the subsystem).  Three important things need to be kept in mind: (i) The
fields coming out of any interaction vertex belongs to the same
replica, (ii) The propagators $\tilde{G}(r,t;r',t')$ are supported over the entire system and generically lack translational invariance due to the presence of the entanglement cut, and (iii) Certain diagrams which vanish in standard Keldysh
field theory give finite contribution, as $\tilde{G}^{\rm\,aa}$ are finite in this
theory. The first order correction to $S^{n}$ is shown in
Fig.~\ref{fig:dblprop}(c). The first of these diagrams (the
direct contribution) evaluates to
\begin{widetext}
	\begin{equation}
		\delta S^{(n)} = (-\ii)\,\sum_\alpha\,\int\dd{t} \sum_{r,r'}
		\left[\tilde{G}^{\rm\, ss}_{\alpha\alpha}(r,t;r,t)+\tilde{G}^{\rm\, aa}_{\alpha\alpha}(r,t;r,t)\right] U(r,r')  \left[\tilde{G}^{\rm\, as}_{\alpha\alpha}(r',t;r',t)+\tilde{G}^{\rm\,  sa}_{\alpha\alpha}(r',t;r',t)\right],
	\end{equation}
	while the second diagram (the exchange contribution) is given by
		\begin{equation}
			\delta S^{(n)} = (\ii)\,\sum_\alpha\,\int\dd{t} \sum_{r,r'}
			\left[\tilde{G}^{\rm\, ss}_{\alpha\alpha}(r,t;r',t)+\tilde{G}^{\rm\, aa}_{\alpha\alpha}(r,t;r',t)\right] U(r,r') \left[\tilde{G}^{\rm\, as}_{\alpha\alpha}(r',t;r,t)+\tilde{G}^{\rm\,  sa}_{\alpha\alpha}(r',t;r,t)\right].
	\end{equation}
\end{widetext}
One can similarly evaluate other diagrams, some of which are shown in
Fig.~\ref{fig:dblprop}(d). One can thus reconstruct the diagrammatic
series in terms of the replica propagators and the original
interaction vertices. We note that while the propagators cannot be
related to anything physical, the number of diagrams reduce
considerably in this way of grouping the terms. However, in absence of
physical correlators, one needs to construct useful approximate
truncations of the diagrams. While a perturbation theory
immediately provides a truncation, one should be more careful about
constructing non-perturbative approximations in this non-standard
Keldysh field theory.

\section{Conclusions}\label{Conclusions}

In this paper, we have formulated a new way of calculating entanglement
entropy of a generic interacting Fermionic system from the knowledge of correlation
functions in the subsystem. Using a Wigner function based method,
coupled with Schwinger Keldysh field theory, we show that the $n$\txup{th}
R\'{e}nyi entropy $S^{(n)}$ is the Keldysh free energy of a theory of $n$
replicas which are coupled by inter-replica currents, which exist in
the subsystem. These currents are local in space-time, i.e. they are
turned on between same degrees of freedom at the time of measurement
of the entanglement theory. These currents have a structure which is
not allowed in usual Keldysh field theory with a single replica, and
hence we do not have an equivalent formulation in usual single-contour
field theories. These currents implement the boundary condition
matching required in standard replica formulation of entanglement
theory. 

Starting from this description of EE as a free
energy in presence of inter-replica currents, we show that the
EE can be written as a sum of terms which require
knowledge of progressively higher order connected correlators in the
system. These correlators are usual field theoretic observables,
calculated in a standard ffield theory with no replicas and correspond
to measureables in the system. We provide an analytic formula for the single-particle
contribution to entanglement; we also provide a diagrammatic
construction for the contribution of higher particle correlators to
EE. We thus relate the correlation
functions in a system to its . These constructions
are agnostic to how the correlators are calculated, and hence forms a
universal basis for further approximate calculations. One can also use
experimentally measured correlation functions in these formulae to
calculate entanglement. They provide estimates for entanglement when only a
few order correlation functions are known. This reduces the complexity
of calculating entanglement entropies vis a vis direct methods which
requires the knowledge of the full quantum many body state.

We have considered how one can implement the analytic continuation
required to obtain the von-Neumann entropy $S$ from the R\'{e}nyi entropy. We
obtain analytic formula for the single-particle contribution to $S$ in
terms of interacting one-particle distributions. We show that a large
class of diagrams for multi-particle contributions vanish under the
analytic continuation. We calculate an analytic formula for the
first non-trivial diagram for two-particle contribution to $S$.

Technically we achieve this in two different ways,(a) by constructing
an expansion around an interacting theory with independent
replicas. This provides a relation between observables and
entanglement and is useful for gaining insights. (b) by constructing
an expansion around a non-interacting theory of coupled
replicas. While this method is less insightful, it provides a simpler
construction of diagrams, since large class of individual diagrams in
the first method are re-summed into single objects in this case. While
the first method provides relation between correlations and
entanglement, the second method provides a relation between vertex
functions and entanglement, and may be more suitable for treatments
like renormalization group analysis.

We note that what we have done here is akin to setting up a general
diagrammatic expansion and writing down the Feynman diagrams and
Feynman rules for the calculation of entanglement
entropy. Calculations for particular systems would require further
approximations. One can ask the following question: In general
correlations will be calculated using approximation methods. One would
have to further truncate/approximate using a subset of the diagrams we
have drawn here. For simple approximations like perturbation theory or
large $N$ approximations,
it is clear how such a truncation will happen. However for
non-perturbative approximations, it may turn out that certain
approximations for correlators are compatible with certain subsets of
these diagrams. Is there a general rule for such compatibility? We do
not take up this question in this work, but leave it as a general
question to be answered in future works.
\begin{acknowledgments}
The authors acknowledge useful discussions with Subir Sachdev, Mohit
Randeria, Gautam Mandal, Sandip Trivedi, Kedar Damle and Onkar Parrikkar. 
SM and RS acknowledge support of the Department of Atomic Energy, Government of India, for support under Project Identification No. RTI 4002.
\end{acknowledgments}
\appendix

\section{Equal time Keldysh correlators in terms of operators}\label{sec:correlators}

In this section we provide the explicit form of the many-particle equal time Keldysh correlators in terms of electron operators. These will be useful when connecting our formalism to numerical simulations or experiments measuring such correlators.

Like in usual quantum field theory, the Schwinger Keldysh partition function in presence of sources, $\mathcal{Z}[J_\mathrm{s,a},\bar{J}_\mathrm{s,a}]$ is the generating function of $k$-particle correlators, and $\ln\mathcal{Z}$ is the generating function for connected correlators. In particular, to get the $k$-particle Keldysh correlator involving all symmetric fields, we take derivatives of $\ln\mathcal{Z}$ w.r.to the antisymmetric sources,
\begin{equation}\label{eq:generatingfn}
	\begin{aligned}
		&\ev{\psi_\mathrm{s}(1)\cdots\psi_\mathrm{s}(k)\pbar_\mathrm{s}(k')\cdots\pbar_\mathrm{s}(1')}_c\\
		&=\left.
		\frac{\delta^{2k}\ln\mathcal{Z}[J_\mathrm{a,s},\bar{J}_\mathrm{a,s}]}%
		{\delta\bar{J}_\mathrm{a}(1)\cdots\delta\bar{J}_\mathrm{a}(k)\,%
			\delta J_\mathrm{a}(k')\cdots\delta J_\mathrm{a}(1')}%
		\right|_{J_\mathrm{a,s},\bar{J}_\mathrm{a,s}=0},
	\end{aligned}
\end{equation}
where the field arguments are shorthand for coordinates, $(1)\equiv(i_1,t_1)$, $(1')\equiv(i_1',t_1')$ etc.
We now use the fact that the Wigner Characteristic function $\chi(\ve{\zbar},\ve{\ze};t_o)$ is a Keldysh partition function in the presence of instantaneous sources\cite{Chakraborty.Sensarma_PRL21_NonequilibriumDynamicsRenyi,moitra_entanglement_2020}. In case of the equal time Keldysh correlator, Eq.~\eqref{eq:sources} allows us to replace $\cal Z$ in Eq.~\eqref{eq:generatingfn} with $\chi(\ve{\zbar},\ve{\ze};t_o)$, and employing Eq.~\eqref{eq:chi_action}, the functional derivatives w.r.to the sources $J_\tx{a}$ simplify to partial derivatives w.r.to the Grassman 
variables $\ve{\ze},\ve{\zbar}$,
%
\begin{equation}
	\begin{aligned}
		\ii^kG^{(k)}_c(i_1\cdot\cdot\, i_k|i'_1\cdot\cdot\, i'_k,t_o)
		=2^k\!\left.
		\frac{\del^{2k}\ln\chi(\ve{\zbar},\ve{\ze};t_o)}%
		{\del\zbar_1\cdot\cdot\,\del\zbar_k\,%
			\del \ze_{k'}\cdot\cdot\,\del \ze_{1'}}%
		\right|_{\substack{\ve{\ze}=0\\\ve{\zbar}=0}}.
	\end{aligned}
\end{equation}
Here $\ze_k$ is shorthand notation for the Grassman variable at site $i_k$, $\ze_{i_k}$.
For the rest of this section we suppress the time label $t_o$ for brevity. 
To illustrate how this expression with partial derivatives simplifies it is convenient to first consider the full
correlation function $G^{(k)}$ generated from $\chi_A$. The resultant expression's connected piece will then reproduce the connected correlation function $G^{(k)}_c$.

From the definition of $\chi$ in Eq.~\eqref{eq:chi} as an expectation value of the Fermionic displacement operator $\opr{D}(\ve{\zbar},\ve{\ze})$, the correlation function $G^{(k)}$ can be written as
\begin{equation}
	\begin{aligned}
		\ii^kG^{(k)}(i_1\cdot\cdot\, i_k|i'_1\cdot\cdot\, i'_k)
		=2^k\!\left.
		\ev{\frac{\del^{2k}\opr{D}(\ve{\zbar},\ve{\ze})}%
			{\del\zbar_1\cdot\cdot\,\del\zbar_k\,%
				\del \ze_{k'}\cdot\cdot\,\del \ze_{1'}}}%
		\right|_{\substack{\ve{\ze}=0\\\ve{\zbar}=0}}.
	\end{aligned}
\end{equation}
We can use the anti-commutation relations amongst the Grassman variables and Fermion operators to get a simplified form for $\opr{D}$~\cite{moitra_entanglement_2020}, 
\begin{align}
	\opr{D}(\ve{\zbar},\ve{\ze})&\equiv\e^{\sum_{i}\cd_i\ze_i-\zbar_i\cc_i}\no\\
	&=\prod_{i}\left[1+\cd_i\ze_i-\zbar_i\cc_i+\ze_i\zbar_i(\cc_i\cd_i-\tfrac{1}{2})\right].
\end{align}
It is then immediate to read off the partial derivatives,
\begin{equation}
	\begin{gathered}
		\left.\frac{\del\opr{D}}{\del\ze_m}\right|_{\ve{\ze},\ve{\zbar}=0}=-\cd_{i_m}%
		\qc
		\left.\frac{\del\opr{D}}{\del\zbar_m}\right|_{\ve{\ze},\ve{\zbar}=0}=-\cc_{i_m},\\[2pt]%
		\tx{and}\quad%
		\left.\frac{\del^2\opr{D}}{\del\zbar_m\del\ze_m}\right|_{\ve{\ze},\ve{\zbar}=0}=\cc_{i_m}\cd_{i_m}-\tfrac{1}{2}.
	\end{gathered}
\end{equation}
In the case where none of the $(m)$ and $(m')$ coordinates coincide, it is easy to see that the $k$-particle correlator is,
\begin{equation}
	\ii^kG^{(k)}(i_1\cdots i_k|i'_1\cdots i'_k)
	=2^k
	\ev{\cc_{i\vphantom{'}_1}\cdots\cc_{i\vphantom{'}_k}\cd_{i'_k}\cdots\cd_{i'_1}},
\end{equation}
which in turn implies that the connected correlator is given by
\begin{equation}\label{eq:connectedGkc}
	\ii^kG^{(k)}_c(i_1\cdots i_k|i'_1\cdots i'_k)
	=(-2)^k
	\ev{\cd_{i'_k}\cdots\cd_{i'_1}\cc_{i\vphantom{'}_1}\cdots\cc_{i\vphantom{'}_k}}_c.
\end{equation}
The extra $(-1)^k$ results from normal ordering the operators. In case of coincident coordinates, the expression for the $k$-particle correlator picks up extra terms with lower order correlators ($k-1$, $k-2$, etc.). However these extra pieces get cancelled in the subtractions to get the connected $k$-particle correlator, making Eq.~\eqref{eq:connectedGkc} valid for arbitrary coordinates.
\section{Analytic continuation of two $G^{(2)}_c$ diagram  }\label{sec:appendix}

The objective of this section is to work out the analytic continuation of a particular 
diagram given in Fig.~\ref{fig:S2p}(a-ii), henceforth referred to as  $\mathfrak{D}$.

For the following it is convenient ot define $\{\ket{i}\}$ as a complete set of states in $A$, with $\ket{i}$ localised on degree of freedom $i$. We can then make the identifications
	\begin{equation*}
		\mel{i}{\mathcal{\opr{V}}_{\alpha\beta}}{j}\equiv\mathcal{V}_{\alpha\beta}(i,j),%
		\quad%
		\mel{ij}{\opr{G}^{(2)}_c}{kl}\equiv G^{(2)}_c(i,j|k,l;t_o).
	\end{equation*}
where $\ket{ij}=\ket{i}\otimes\ket{j}$. It is then immediate to rewrite Eq.~\eqref{eq:ladder} in more compact notation as
\begin{equation}\label{eq:twog2diagram}
	\mathfrak{D}=\frac{1}{2}\sum_{\alpha,\beta=1}^{n}%
	\Tr_{A}[%
	\mathcal{\opr{V}}_{\alpha\beta}\otimes\mathcal{\opr{V}}_{\alpha\beta}\opr{G}^{(2)}_c%
	\,\mathcal{\opr{V}}_{\beta\alpha}\otimes\mathcal{\opr{V}}_{\beta\alpha}\opr{G}^{(2)}_c%
	],
\end{equation}
where $\Tr_A$ is now understood to be over two copies of $A$.
From Eq.~\eqref{eq:Vmat} it is clear that the matrix $[\mathcal{\opr{V}}_{\alpha\beta}\otimes\mathcal{\opr{V}}_{\alpha\beta}]$ is block circulant in replica indices, and hence the sum over the same in Eq.~\eqref{eq:twog2diagram} can be simplified to read
	\begin{equation}\label{eq:trblocks}
		\begin{aligned}
		\mathfrak{D}&=\frac{n}{2}\Tr_A[\opr{v}_0^{\otimes2}\,\opr{G}^{(2)}_c\,\opr{v}_0^{\otimes2}\opr{G}^{(2)}_c]\\%
		&\hphantom{=}+\frac{n}{2}\sum_{k=1}^{n-1}%
		\Tr_A[\opr{v}_k^{\otimes2}\,\opr{G}^{(2)}_c\,\opr{v}_{n-k}^{\otimes2}\opr{G}^{(2)}_c],
		\end{aligned}
	\end{equation}
where $\opr{v}_k$ are as defined in Eq.~\eqref{eq:smallv} and $\opr{v}_k^{\otimes2}=\opr{v}_k\otimes\opr{v}_k$. We immediately note that the 
first term in the sum will not contribute in the $n\to1$ limit since $\opr{v}_0\sim(n-1)$ from Eq.~\eqref{eq:v0scaling}. 
Evaluating the rest of the sum is not a priori straightforward as $\opr{v}_k^{\otimes2}$ and $\opr{G}^{(2)}_c$ do not commute in general. 
It is convenient to switch to the basis in which $\opr{v}_k$ is diagonal, namely the eigenbasis of the $\opr{C}$ operator (correlation matrix restricted to the subsystem $A$), $\{\ket{c}\}$ defined as
$\opr{C}\ket{c}=c\ket{c}$.
In this basis, $\opr{v}_k^{\otimes2}$ takes the form,
	\begin{equation}
		\opr{v}_k^{\otimes2}=\sum_{c_1,c_2}%
		\frac{(c_1c_2)^{n-k-1}[(1-c_1)(1-c_2)]^{k-1}}{4[c_1^n+(1-c_1)^n][c_2^n+(1-c_2)^n]}	
		\op{c_1c_2}{c_1c_2}.
	\end{equation}
Here $c_1$ and $c_2$ are eigenvalues of $\opr{C}$, each running over the entire spectrum of the same.
Substituting this form into Eq.~\eqref{eq:trblocks} and ignoring the leading piece with $\opr{v}_0$, we get
		\begin{multline}\no
			\mathfrak{D}=%
			\frac{n}{32}\sum_{\substack{c_1,c_2\\c_3,c_4}}%
			\frac{%
				\mel{c_1c_2}{\opr{G}^{(2)}_c}{c_3c_4}\mel{c_3c_4}{\opr{G}^{(2)}_c}{c_1c_2}
			}{\prod_{j=1}^{4}[c_j^n+(1-c_j)^n]}\\%
			\sum_{k=1}^{n-1}x^{n-k-1}y^{k-1},
		\end{multline}
where we have defined
	\begin{equation}
			x=c_1\,c_2\,(1-c_3)(1-c_4),\quad
			y=(1-c_1)(1-c_2)\,c_3\,c_4.
	\end{equation}
The last sum over replica blocks can now be done trivially,
	\begin{equation*}
		\sum_{k=1}^{n-1}x^{n-k-1}y^{k-1}=\frac{x^{n-1}-y^{n-1}}{x-y}.
	\end{equation*}
To analytically continue the contribution of this diagram to $S$, we look at $\lim_{n\to1}\mathfrak{D}/(1-n)$, which gives back the result quoted in Eq.~\eqref{eq:svn_twog2}.
\bibliography{fermionS2,EEgeneral,2Dfermionentanglement,fermiliquids,mathresources,ownpapers}
\end{document}